\newif\ifPDF
\ifx\pdfoutput\undefined\PDFfalse
\else\ifnum\pdfoutput > 0\PDFtrue
     \else\PDFfalse
     \fi
\fi
\documentclass[usenatbib]{mn2e}
\ifPDF
  \usepackage[T1]{fontenc}
  \usepackage{aeguill}
  \usepackage[pdftex]{graphicx,color}
  \usepackage{subfigure}
  \usepackage{afterpage}
\else
  \usepackage[T1]{fontenc}
  \usepackage[dvips]{graphicx,color}
  \usepackage{subfigure}
  \usepackage{afterpage}
\fi
\newcommand{\kms}{{km$\;$s$^{-1}$}}

\title[SMC eclipsing binaries]
{Forty eclipsing binaries in the Small Magellanic Cloud:\\ fundamental
parameters and Cloud distance}
\author[R.W. Hilditch, I.D. Howarth \& T.J. Harries]
{R.W.Hilditch$^{1}$, I.D.Howarth$^{2}$ and T.J.Harries$^{3}$\\
$^{1}$School of Physics \& Astronomy, University of St Andrews, North Haugh, St Andrews, KY16 9SS\\
$^{2}$Department of Physics \& Astronomy, University College London, Gower Street, London, WC1E 6BT \\
$^{3}$Department of Physics, University of Exeter, Stocker Road, Exeter, EX4 4QL}
\begin{document}

\date{Received: }

\maketitle

\begin{abstract}
We have conducted a programme to determine the fundamental parameters
of a substantial number of eclipsing binaries of spectral types O~and
B in the Small Magellanic Cloud (SMC). New spectroscopic data,
obtained with the two-degree-field (2dF) multi-object spectrograph on
the 3.9-m Anglo-Australian Telescope, have been used in conjunction
with photometry from the Optical Gravitational Lens Experiment
(OGLE-II) database of SMC eclipsing binaries.  Previously we reported
results for 10 systems; in this second and concluding paper we present
spectral types, masses, radii, temperatures, surface gravities and
luminosities for the components of a further 40 binaries.  The
uncertainties are typically $\pm10\%$ on masses, $\pm4\%$ on radii and
$\pm0.07$ on $\log{L}$.  The full sample of 50 OB-type eclipsing
systems is the largest single set of fundamental parameters
determined for high-mass binaries in any galaxy.  We find that 21 of
the systems studied are in detached configurations, 28 are in
semi-detached post-mass-transfer states, and one is a contact binary.

The overall properties of the detached systems are consistent with
theoretical models for the evolution of single stars with SMC metal
abundances ($Z \simeq 0.004$); in particular, observed and
evolutionary masses are in excellent agreement. Although there are no
{\em directly} applicable published models, the overall properties of the
semi-detached systems are consistent with them being in the slow phase
of mass transfer in case~A. About 40\%\ of these semi-detached systems
show photometric evidence of orbital-phase-dependent absorption by a
gas stream falling from the inner Lagrangian point on the secondary
star towards the primary star. This sample demonstrates that case-A
mass transfer is a common occurrence amongst high-mass binaries with
initial orbital periods $P\la5$ days, and that this slow phase has a
comparable duration to the detached phase preceding it.

Each system provides a primary distance indicator. We find a mean
distance modulus to the SMC of $18.91\pm0.03\pm0.1$ (internal and
external uncertainties; $D = 60.6\pm1.0$~kpc). This value represents
one of the most precise available determinations of the distance to
the SMC.
\end{abstract}


\section{Introduction}

In a previous paper, Harries, Hilditch \& Howarth (2003, hereinafter
H$^3$03) reported the first results of an observational programme to
determine the fundamental para\-meters of stars in eclipsing binaries in
the Small Magellanic Cloud (SMC). The rationale for this programme was
fully described in H$^3$03, so it suffices to state here that the twin
aims of this investigation are to provide tests of evolutionary models
for high-mass binary stars, including those that undergo
mass-exchanging interactions at early stages in their evolution; and
to provide a set of primary distance indicators to determine the
distance to, and line-of-sight depth of, the SMC, independently of all
other distance calibrators.  Since publication of H$^3$03, a Joint
Discussion on `Extragalactic Binaries' was held at the 2003 IAU
General Assembly, and the papers presented there by many authors
provide a broad overview of recent progress in this field (cf.\ Ribas \&
Gim\'{e}nez 2004).

H$^3$03 reported combined spectroscopic and photometric analyses of 10
relatively bright binaries; we now present follow-up observations and
analyses of 40 additional targets.  With such a large sample we are
insensitive to anomalous individual results (whether they arise from
astrophysical effects or from problems with data, such as unresolved
blending in the photometry), as well as gaining from simple
statistical leverage that gives our results (and particularly our
distance determination) very high weight.  Furthermore, by introducing
radial-velocity data, albeit of modest quality, we gain enormously
over analyses based solely on light-curve solutions.

We summarize our observing programme in Section~\ref{SpObs}, and the
spectroscopic and light-curve analyses in Section~\ref{DatAnal}.
Section~\ref{Params} describes comparisons between the derived
fundamental astrophysical parameters and evolutionary models for
single stars and interacting binary stars, while Section~\ref{Dists}
reviews distance determinations.

\section[]{Spectroscopic Observations}
\label{SpObs}

Our programme was devised in 2000, aiming to exploit both the
publication of the Optical Gravitational Lens Experiment photometric
survey (OGLE-II; Udalski {\em et al.}\ 1998) and the commissioning of
the two-degree-field (2dF) multi-object spectrograph on the 3.9-m
Anglo-Australian Telescope (AAT).  Extensive simulations for realistic
expectations of time allocations showed that satisfactory orbital
solutions should be obtainable for systems with $B\la16$ mag and
orbital periods $P\la5$ days -- constraints ensuring both adequate
phase coverage for a reasonable sample, and sufficient signal-to-noise
ratio on the faintest targets (target $S/N\ga25$ per wavelength sample
per observation).  With these boundary conditions (and some further
limitations introduced by practical constraints on the positioning of
the 2dF optical fibres) we identified an initial sample of 169 SMC
eclipsing binaries.

Our spectroscopy of these targets was obtained in two overlapping
2$^\circ$-diameter fields in the central regions of the SMC; details
of the field centres, and checks on the astrometry, are described in
H$^3$03.  The data were taken using 2dF, in service mode (as is
standard for this instrument).  Each full observation consisted of two
1800-s integrations (on, typically, 100 targets simultaneously), plus
arc and flat-field frames.  Integration times of 1800s were used for
the target exposures, in seeing of typically 1.5--2$''$ (cp.\ fibre
diameters of 2$''$). We used the marginally better-performing of the
two spectrographs, with a 1200B grating providing a mean reciprocal
dispersion of 1.1\AA\,px$^{-1}$ and a resolution of 2\AA\ (the best
available with this instrument). The wavelength range of the data is
3855--4910\AA, which ensures excellent coverage of the H, He$\;${\sc
i} and He$\;${\sc ii} lines that dominate the blue spectra of O and
early B stars. The 2{\sc d}FDR package (Lewis et al.~2002) was used
with default settings to reduce all the data.  The resultant spectra
have $S/N$ values ranging from $\sim20$ to $\sim50$.

From allocations of 12 nights over the 2001--2003 SMC observing
seasons, worthwhile observations were secured on only $5\frac{1}{2}$
nights.  In consequence, the orbital phase coverage is generally
poorer than planned, and for any given target depends on the orbital
ephemeris and location (in the field overlap region or otherwise).
The final database has a total of more than 5200 spectra of 169
systems, with from 5 to 55 observations per target. Of these 169
systems, $\sim$40\%\ do not have observations at or near quadrature
phases (even in some cases with 55 spectra!).  The remaining $\sim$100
stars were studied in greater detail, using the procedures described
in Section~\ref{DatAnal}.

\section[]{Data Analysis}
\label{DatAnal}

The spectra were normalized by using an interactively defined spline
fit to line-free continuum regions. (In examining each spectrum, any
remaining noise spikes caused by cosmic rays etc. could be removed.)
The spectra were then velocity-corrected to the heliocentric reference
frame, and `windowed' to isolate specific regions which include strong
absorption lines, such as 4300--4500\AA\ (H$\gamma$, He$\;${\sc i}
$\lambda$4388, $\lambda$4471).  Orbital phases were calculated from
photometric
ephemerides\footnote{In consequence, when we refer to the `primary
component' we mean the star which is at superior conjunction at
photometric phase zero (i.e., eclipsed at primary eclipse), which
is the component with greater surface brightness -- not necessarily
the more massive component.}
published by Wyrzykowski et al. (2004, based on 4 seasons
of OGLE-II photometry), where available; otherwise we adopted results
from Udalski et al. (1998; $1\frac{1}{2}$
seasons).

\subsection{Orbital parameters and component spectra}

H$^3$03 described in detail the implementation of spectral
disentangling codes which permit precise and direct
determinations of the orbital parameters of the binary components,
without the need for intermediate steps (i.e., measurement of velocities
for each component for every observation).  We demonstrated that our
codes, which implement the algorithm described by Simon \& Sturm
(1994), give results that agree well with orbital solutions based on
the cross-correlation methods previously more frequently used for
measuring radial velocities.
Compared to these `traditional' methods, spectral disentangling not only
yields orbital solutions of better precision (or, indeed, orbital
solutions from data that would not be amenable to cross-correlation
analyses), but also gives disentangled average spectra for each
binary component (at the expense of not returning `per observation'
velocity measurements).  Some excellent review papers on the successes
of spectral disentangling procedures are given by Hadrava (2004),
Hensberge \& Pavlovski (2004), Holmgren (2004) and Ilijic (2004).

\subsubsection{Velocities}

For each system we searched for the best (least-squares) orbital
solution by conducting grid searches in orbital semi-amplitudes ($K_{1},
K_{2}$), given the period and reference
epoch.  Circular orbits were assumed unless the light curve
demonstrated eccentricity.\footnote{In most cases, the light curves
show no evidence of departures from circular orbits, with secondary
eclipses at phase 0.50, and both eclipses of equal durations. In a few
cases (discussed in Section~\ref{Notes}) the orbits are, however,
detectably eccentric, with secondary eclipses displaced from phase
0.50 by up to 0.10.}  It was found that using data in the phase ranges
0.20--0.30 and 0.70--0.80 gave the most reliable and consistent
results; as necessary, wider phase ranges were employed (0.15--0.35
and 0.65--0.85), but observations closer to conjunctions offer no
useful leverage on orbital-velocity amplitudes.

At our spectral resolution, many binaries are clearly double-lined at
quadrature phases, and well-defined orbital solutions are then readily
obtained. However, the disentangling procedure is somewhat sensitive
to the combination of orbital phase coverage and $S/N$ in the data --
important limiting factors in our sample.  Where the grid search
yielded multiple $\chi^{2}$ minima, the consequent ambiguity could
sometimes be safely resolved by further investigation (identifying prefered
solutions on physical grounds).  In all cases, sequentially finer grid
searches were used to refine the solution.  

The outcome of this survey was that, of the $\sim$100 systems examined
in detail, 40 yielded reliable orbital parameters (in addition to the
10 systems already discussed in H$^3$03).  The remainder did not
provide sensible solutions, in that derived masses and distances were
obviously discrepant (typically by factors of 2 or more).
Spectroscopic solutions for the 40 `good' systems are given in
Table~\ref{circorb} for circular orbits, and Table~\ref{eccorb} for
eccentric orbits.  As the disentangling procedure returns only the
semi-amplitudes, the tabulated systemic velocities were determined
separately, from a cross-correlation analysis of the disentangled mean
spectra against a template spectrum of the O9$\;$V star 10~Lac, using
the \textsc{vcross} code (Hill 1982).  The values reported for each
binary are the flux-weighted means of the two components.  For
eccentric orbits, values of eccentricity, $e$, and longitude of
periastron, $\omega$, were determined from the light-curve solutions
(from the position of secondary eclipse and the durations of the two
eclipses), and then held fixed in the spectroscopic solutions (because
of selective phase coverage of the spectroscopy).\footnote{The
light-curve solutions require input (the mass ratio) from the
spectroscopic orbit, and the spectroscopic solutions require input
(`non-keplerian' corrections, as well as $e$ and $\omega$) from the
light-curve solutions.  The velocity corrections rarely exceed 5~\kms\
and a single iteration between light-curve and spectroscopic solutions
gave convergence (including any adjustments to $e$ and $\omega$).}

The published velocity semi-amplitudes of the 10 systems studied by
H$^3$03 were derived from spectra obtained only in the first, 2001,
observing session.  Since the numbers of useable spectra of these 10
systems has subsequently increased (and the original data have been
re-reduced), new orbital solutions were conducted.  The mean
difference between the published and new semi-amplitudes is
$-2\pm4$~\kms\ (s.e.), and in no case did the original and new
semi-amplitudes differ by more than the joint 1-$\sigma$
uncertainties.  The mean difference in systemic velocities is
$-11\pm4$~\kms\ (s.e., H$^3$03 minus this paper).  This difference
reflects the fact that in H$^3$03, the heliocentric velocity
corrections ($\sim{6}$~\kms) were inadvertently applied with the
incorrect sign, such that the previously published $\gamma$~velocities
should be corrected by +12~\kms. It is pleasing that this known
mistake should be recovered in the data analysis, and that the
expected uncertainties in the velocities ($\sim\pm15$~\kms) should be
consistent with the dispersion in the results.  Other than the
correction to the $\gamma$~velocities, there is clearly no need to
replace the H$^3$03 orbital solutions.

\begin{table*}
\caption[]{Circular-orbit solutions (including `non-keplerian' corrections).  
Identifiers are from
Udalski {\em et al.}\ (1998, PSF) and
Wyrzykowski et al. (2004, DIA),
and
for each target there are $n$
useful spectra, characterized by the stated $S/N$.}
\label{circorb}
\begin{tabular}{rcrc rcr rcr rcr}
\hline
\multicolumn{1}{c}{OGLE-PSF}&
\multicolumn{1}{c}{OGLE-DIA}&
\multicolumn{1}{c}{$n$}&
\multicolumn{1}{c}{$S/N$}&
\multicolumn{3}{c}{$\qquad K_{1}$}&
\multicolumn{3}{c}{$\qquad K_{2}$}&
\multicolumn{3}{c}{$\qquad \gamma$} \\
\multicolumn{1}{c}{identifier}&
\multicolumn{1}{c}{identifier}&
&
&
\multicolumn{3}{c}{\qquad (\kms)}&
\multicolumn{3}{c}{\qquad (\kms)}&
\multicolumn{3}{c}{\qquad (\kms)} \\
\hline
 1~099121&003851.98$-$733433.2&$11$&$25$&$\qquad 152$&$\pm$&$11$&$\qquad 259$&$\pm$&$6$&$\qquad 182$&$\pm$&$15$ \\
 4~056804&004633.14$-$732217.0&$4$&$25$&$307$&$\pm$&$5$&$280$&$\pm$&$2$&$155$&$\pm$&$5$ \\
 4~103706&004725.50$-$732716.7&$7$&$40$&$203$&$\pm$&$6$&$359$&$\pm$&$4$&$205$&$\pm$&$13$ \\
 4~110409&004700.19$-$731843.1&$8$&$25$&$160$&$\pm$&$11$&$247$&$\pm$&$3$&$204$&$\pm$&$18$ \\
 4~163552&004753.24$-$731556.5&$5$&$30$&$257$&$\pm$&$11$&$275$&$\pm$&$9$&$170$&$\pm$&$3$ \\
 5~026631&004859.84$-$731328.8&$10$&$25$&$234$&$\pm$&$8$&$239$&$\pm$&$5$&$167$&$\pm$&$6$ \\
 5~060548&004835.40$-$725256.5&$22$&$40$&$151$&$\pm$&$4$&$187$&$\pm$&$3$&$179$&$\pm$&$5$ \\
 5~095194&004950.42$-$731931.6&$12$&$40$&$247$&$\pm$&$23$&$215$&$\pm$&$22$&$193$&$\pm$&$6$ \\
 5~140701&004943.08$-$725109.0&$12$&$30$&$137$&$\pm$&$9$&$179$&$\pm$&$7$&$181$&$\pm$&$9$ \\
 5~180064&005044.74$-$731739.9&$15$&$25$&$146$&$\pm$&$3$&$222$&$\pm$&$3$&$153$&$\pm$&$6$ \\
 5~208049&005045.00$-$725844.4&$7$&$25$&$108$&$\pm$&$3$&$226$&$\pm$&$1$&$170$&$\pm$&$4$ \\
 5~243188&005118.78$-$733015.8&$18$&$40$&$247$&$\pm$&$14$&$362$&$\pm$&$5$&$206$&$\pm$&$5$ \\
 5~255984&005129.62$-$732137.7&$7$&$25$&$176$&$\pm$&$38$&$292$&$\pm$&$8$&$209$&$\pm$&$3$ \\
 5~277080&005111.68$-$730520.3&$13$&$25$&$200$&$\pm$&$10$&$308$&$\pm$&$5$&$199$&$\pm$&$23$ \\
 5~300549&005123.57$-$725224.1&$9$&$40$&$253$&$\pm$&$9$&$369$&$\pm$&$4$&$196$&$\pm$&$5$ \\
 5~305884&005120.17$-$724942.9&$17$&$50$&$232$&$\pm$&$14$&$252$&$\pm$&$5$&$184$&$\pm$&$4$ \\
 6~011141&005203.96$-$731849.3&$13$&$25$&$263$&$\pm$&$4$&$281$&$\pm$&$2$&$222$&$\pm$&$11$ \\
 6~152981&005241.88$-$724622.4&$6$&$30$&$180$&$\pm$&$4$&$276$&$\pm$&$3$&$212$&$\pm$&$9$ \\
 6~180084&005342.42$-$732319.9&$8$&$30$&$261$&$\pm$&$22$&$305$&$\pm$&$4$&$191$&$\pm$&$4$ \\
 6~251047&005344.05$-$723124.0&$15$&$25$&$150$&$\pm$&$3$&$221$&$\pm$&$2$&$193$&$\pm$&$5$ \\
 6~311225&005402.00$-$724221.6&$20$&$25$&$176$&$\pm$&$3$&$312$&$\pm$&$2$&$200$&$\pm$&$4$ \\
 6~319960&005405.26$-$723426.0&$15$&$40$&$130$&$\pm$&$11$&$204$&$\pm$&$4$&$165$&$\pm$&$20$ \\
 7~066175&005438.22$-$723206.2&$20$&$45$&$155$&$\pm$&$14$&$264$&$\pm$&$7$&$212$&$\pm$&$3$ \\
 7~120044&005531.57$-$724307.8&$12$&$25$&$257$&$\pm$&$6$&$258$&$\pm$&$4$&$192$&$\pm$&$6$ \\
 7~142073&005554.44$-$722808.5&$6$&$30$&$127$&$\pm$&$10$&$254$&$\pm$&$9$&$193$&$\pm$&$7$ \\
 7~189660&005637.30$-$724143.3&$10$&$20$&$207$&$\pm$&$10$&$310$&$\pm$&$10$&$197$&$\pm$&$2$ \\
 7~193779&005621.80$-$723701.7&$21$&$20$&$147$&$\pm$&$13$&$289$&$\pm$&$8$&$207$&$\pm$&$3$ \\
 8~087175&005830.98$-$723913.8&$7$&$20$&$268$&$\pm$&$21$&$300$&$\pm$&$6$&$233$&$\pm$&$10$ \\
 8~104222&005825.08$-$721909.8&$11$&$25$&$257$&$\pm$&$14$&$277$&$\pm$&$6$&$250$&$\pm$&$2$ \\
 8~209964&010016.05$-$721243.9&$14$&$50$&$180$&$\pm$&$6$&$234$&$\pm$&$4$&$212$&$\pm$&$18$ \\
 9~010098&010052.90$-$724748.6&$7$&$30$&$260$&$\pm$&$23$&$339$&$\pm$&$11$&$216$&$\pm$&$8$ \\
 9~047454&010052.04$-$720705.5&$6$&$25$&$199$&$\pm$&$21$&$271$&$\pm$&$20$&$187$&$\pm$&$12$ \\
 9~064498&010117.26$-$724232.1&$10$&$20$&$76$&$\pm$&$9$&$235$&$\pm$&$5$&$206$&$\pm$&$7$ \\
10~094559&   $\cdots$         &$8$&$30$&$221$&$\pm$&$16$&$267$&$\pm$&$7$&$135$&$\pm$&$7$ \\
10~108086&   $\cdots$         &$6$&$40$&$317$&$\pm$&$20$&$375$&$\pm$&$7$&$226$&$\pm$&$15$ \\
10~110440&   $\cdots$         &$12$&$25$&$160$&$\pm$&$36$&$292$&$\pm$&$13$&$275$&$\pm$&$12$ \\
\hline
\end{tabular}
\end{table*}

\begin{table*}
\caption[]{Eccentric-orbit solutions; $\omega$ is the longitude of periastron
measured from the ascending node for
the primary component.
Semi-amplitudes include `non-keplerian' corrections.
Identifiers are from
Wyrzykowski et al. (2004, DIA) and
Udalski {\em et al.}\ (1998, PSF).        }
\label{eccorb}
\begin{tabular}{rcccccccc}
\hline
\multicolumn{1}{c}{OGLE-PSF}&
\multicolumn{1}{c}{OGLE-DIA}&
\multicolumn{1}{c}{$n$}&
\multicolumn{1}{c}{$S/N$}&
\multicolumn{1}{c}{$K_{1}$}&
\multicolumn{1}{c}{$K_{2}$}&
\multicolumn{1}{c}{$\gamma$}&
\multicolumn{1}{c}{$e$}&
\multicolumn{1}{c}{$\omega$} \\
\multicolumn{1}{c}{identifier}&
\multicolumn{1}{c}{identifier}&
&
&
\multicolumn{1}{c}{(\kms)}&
\multicolumn{1}{c}{(\kms)}&
\multicolumn{1}{c}{(\kms)}&
&
\multicolumn{1}{c}{($^\circ$)} \\
\hline
  5~311566  &005134.85$-$724545.9 & $15$&$20$&$180$&$227$&$179$&$0.063$&$3.0$ \\
 $ $        &                 $ $ & $ $&$ $&$\pm7$&$\pm3$&$\pm2$&$\pm0.002$&$\pm20.0$ \\
  6~221543  &005340.40$-$725222.0 & $22$&$25$&$200$&$205$&$186$&$0.113$&$316.0$ \\
 $ $        &                 $ $ & $ $&$ $&$\pm1$&$\pm21$&$\pm11$&$\pm0.004$&$\pm1.9$ \\
  7~255621  &$\cdots$             & $11$&$20$&$147$&$190$&$181$&$0.190$&$218.2$ \\
 $ $&$ $    &                 $ $ & $ $&$\pm6$&$\pm7$&$\pm11$&$\pm0.006$&$\pm2.9$ \\
  10~037156 &$\cdots$             & $14$&$30$&$231$&$265$&$239$&$0.013$&$202.0$ \\
 $ $        &                 $ $ & $ $$\pm3$&$\pm2$&$\pm10$&$\pm0.009$&$\pm27.0$ \\

\hline
\end{tabular}
\end{table*}

\subsubsection{Spectral types}

The separated component spectra that result from the disentangling
procedure have $S/N$ values that are $\sim\sqrt{n/2}$ greater than for
each of the $n$ individual observed spectra used in the analysis.  We
constructed disentangled spectra for each system by using all
observations with orbital phases 0.15--0.35 and 0.65--0.85, and the
full observed wavelength range.

\begin{figure*}
\includegraphics[width=180mm]{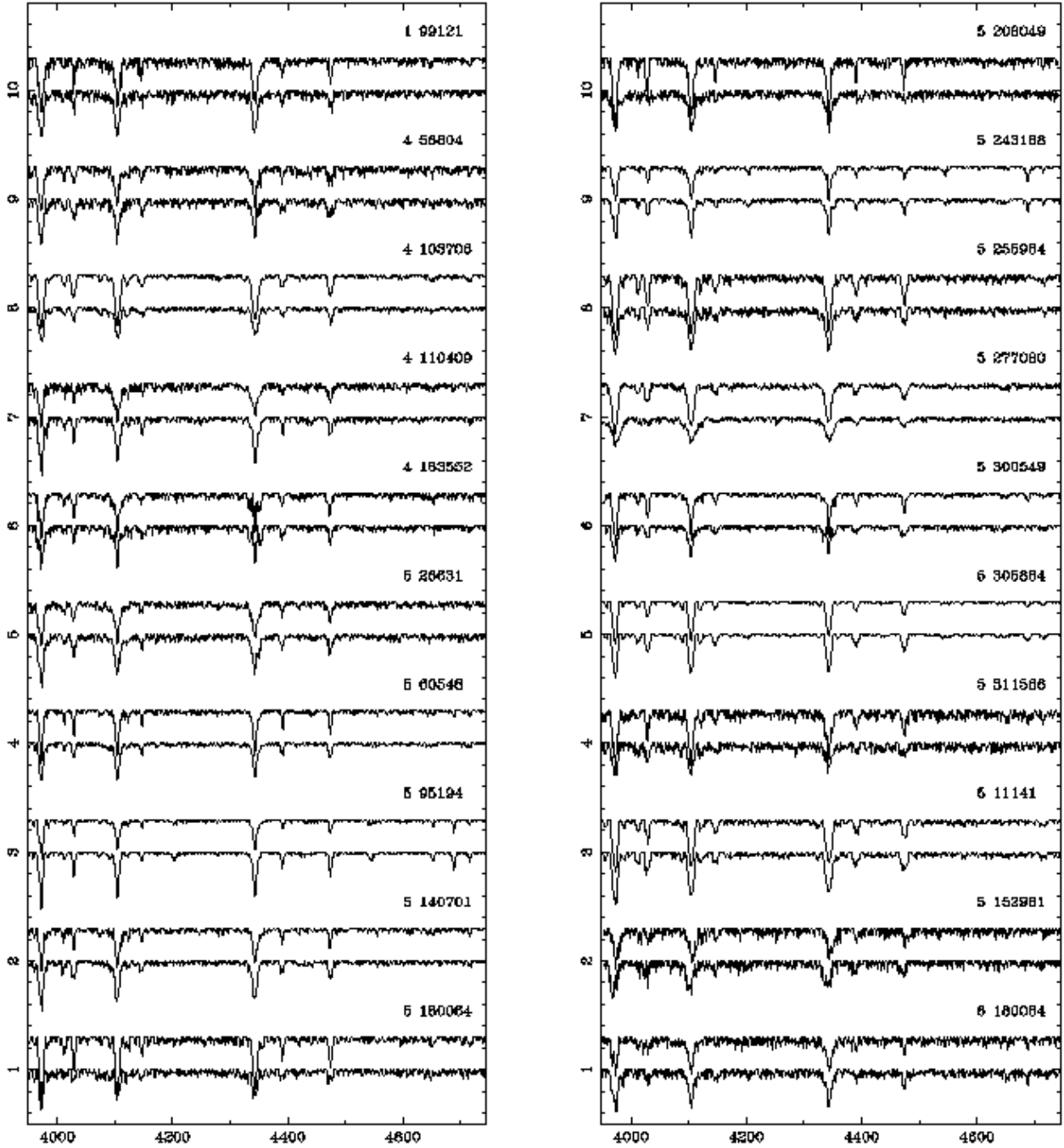} 
 \caption{Disentangled component spectra for 20 binaries (results for
the remainder of the sample are shown in Fig.~\ref{spectra2}).  The
normalized spectra are plotted as relative flux, with the secondary
offset by $-0.5$ continuum units for clarity.  Wavelengths are in \AA;
the strong absorption lines include H$\epsilon$, H$\delta$, H$\gamma$,
and He$\;${\sc i}~$\lambda\lambda$4026, 4144, 4388, 4471, 4713, as
well as He$\;${\sc ii}~$\lambda\lambda$4200, 4542, 4686 in some cases.
}
\label{spectra1}
\end{figure*}

\begin{figure*}
\includegraphics[width=180mm]{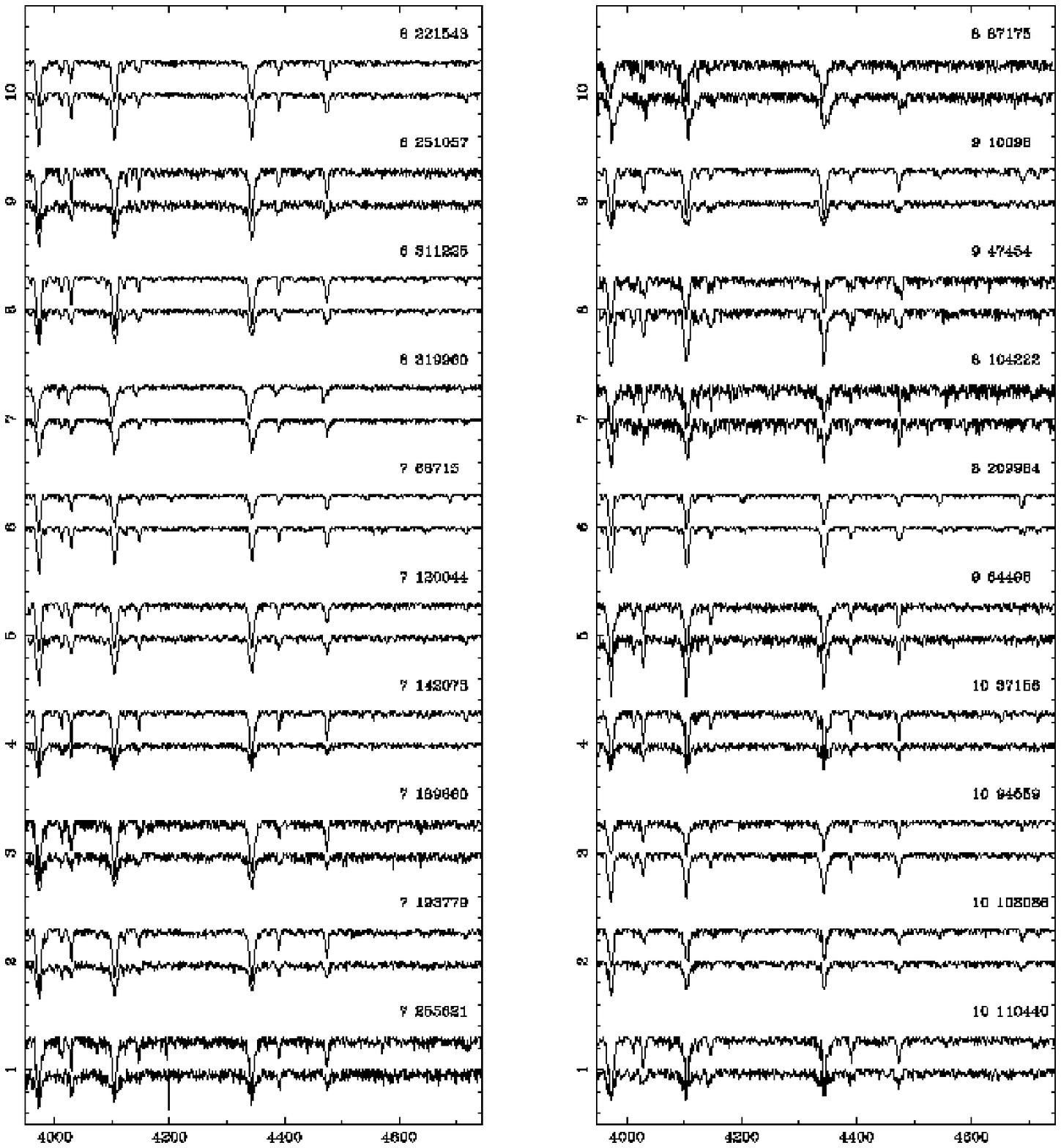} 
 \caption{Disentangled component spectra for 20 binaries (results for
the remainder of the sample are shown in Fig.~\ref{spectra1}).  The
normalized spectra are plotted as relative flux, with the secondary
offset by $-$0.5 continuum units for clarity.  Wavelengths are in \AA;
the strong absorption lines include H$\epsilon$, H$\delta$, H$\gamma$,
and He$\;${\sc i}~$\lambda\lambda$4026, 4144, 4388, 4471, 4713, as
well as He$\;${\sc ii}~$\lambda\lambda$4200, 4542, 4686 in some
cases.}  \label{spectra2}
\end{figure*}

These disentangled component spectra, shown in Figs~\ref{spectra1}
and~\ref{spectra2}, were classified by reference to the digital
spectral atlas of Galactic O,B stars by Walborn \& Fitzpatrick (1990)
and to Lennon's (1997) SMC B-supergiant study.  This task was
performed independently by all three authors in order to assess the
uncertainties associated with the spectral types. Typically we found
that our classifications agreed to within one temperature subtype, and
the adopted modal values for the primary components are given in
Table~\ref{lcsol}. For the (normally weaker) spectra of the secondary
components, we provide the range of spectral types assigned by the
three authors.  We found that the O--B1 stars were quite easy to
classify, but that spectral classifications were more difficult for
later B~types due to the absence of He$\;${\sc ii} lines and, in these
low-metallicity stars, the lack of detectable Si$\;${\sc ii} and other
metal lines (see the discussion by Evans {\em et al.}\ 2004).
However, our classifications of primary components all agree to within
one subtype, and it is these classifications that establish the mean
effective temperatures of the primary components (and hence the system
absolute magnitudes; Section~\ref{Dists}).  In this sense the spectral
types assigned to the secondary components are much less important,
although they do serve as consistency checks for the effective
temperatures of the secondary components.

\subsection{Light-curve analyses }
\label{lcanal}

The  light-curve synthesis code \textsc{light2} (Hill 1979; Hill \&
Rucinski 1993) was used for all analyses of the OGLE-II light
curves. 

\subsubsection{Data sources}

Our initial modeling made use of $I$-band photometry published by
Udalski {\em et al.}\ (1998), based on $1\frac{1}{2}$ seasons of OGLE
observations reduced using standard techniques (point-spread-function
fits to stellar images).  This `OGLE-PSF' database afforded typically
$\sim140$ observations per light curve, with an average photometric
error of 0.014 mag quoted by Udalski {\em et al.}\ (1998).  The
overall rms scatter of the observations about the model fits was 0.019
mag, ranging 0.011--0.028 mag for individual systems, but some of the
OGLE-PSF light curves were not particularly well sampled, with
some eclipses defined by as few as 10 observations.

Wyrzykowski et al (2004) subsequently published results from four
seasons of OGLE $I$-band photometry, reduced by using the
`difference image analysis' (DIA) technique described by Zebrun {\em et
al.} (2001).  Not all our sample stars are included in this OGLE-DIA
database, but for those that are there is improved accuracy in the
ephemerides (though most changes are quite small -- $\sim1$--2
parts in $10^{5}$ for $P_{\rm orb}$); the light curves are better
sampled (typically $\sim300$ observations per target); and the
re-reduced photometry is more accurate (quoted average photometric
error 0.011 mag), at least for uncrowded sources.

We therefore repeated the entire analysis for the 35 systems in our
sample that have DIA photometry, all of which are reported to be
`uncrowded' sources (Wyrzykowski {\em et al.}\ 2004.)  In general, we
found only small changes in the system parameters, but improved
accuracy; the overall rms scatter of the DIA observations about the
model fits is 0.014 mag, ranging 0.005--0.025 mag for individual
systems.  (One exception, 5~311566, is discussed in
Section~\ref{X311566}.) The five systems for which we have only the
PSF photometry still have quite well-defined light curves, and
satisfactory solutions.

\subsubsection{Methodology}

To carry out the light-curve solutions,
mass ratios were fixed from the spectroscopic solutions,
and standard values were used for limb-darkening coefficients
(appropriate to the local temperatures at each grid point over the
surfaces of both stars in each binary), for gravity-darkening
exponents (for stars with radiative envelopes), and for albedos and
electron-scattering fractions (from Hill \& Hutchings 1971).
Synchronous rotation was assumed in all cases.

We solved each light curve for the two stellar radii, the orbital
inclination, and the temperature of the secondary component (strictly
the $I$-band flux ratio between the two stars, but characterized by
black-body fluxes at the appropriate temperatures and effective
wavelengths). The temperature of the primary component was fixed at
the value corresponding to its assigned spectral type, and the default
starting assumption for each solution was that each system is composed
of stars that are detached from their respective Roche lobes.
Semi-detached configurations, with the secondary star filling its
Roche lobe, were adopted only when the detached solutions indicated a
secondary-star volume of 98--100\%\ of its Roche lobe.

The disentangled spectra offer an independent constraint on the flux
ratio: requiring the overall relative strengths of the components'
line spectra to match expectations based on spectral type gives an
estimate of the continuum flux ratio (in the $B$~band).  Because this
is a relatively blunt tool, and because the binaries studied in this
paper are fainter than those discussed in H$^3$03 (and consequently
their spectra have lower $S/N$ values), we did not impose these
spectroscopic flux ratios on the light-curve solutions, but 
used them, where necessary, to distinguish between competing, and
sometimes statistically indistinguishable, solutions to the $I$-band
light curves with detached or semi-detached
configurations.\footnote{For a given system, detached solutions almost
invariably imply flux ratios that are larger than those of
semi-detached solutions, since the differences in radii of the two
stars are generally larger.}

\begin{figure*}
\includegraphics[width=170mm]{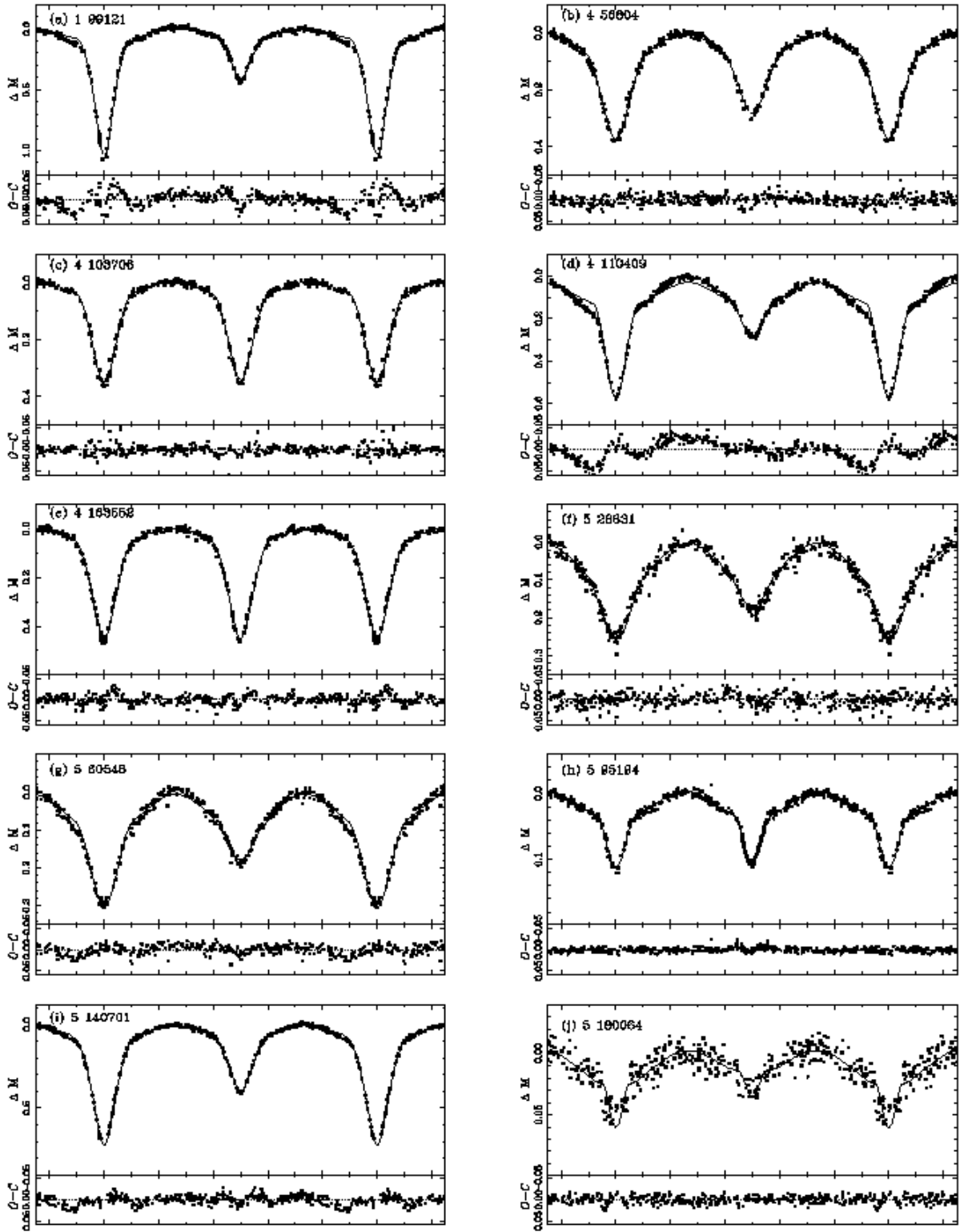} 
 \caption{$I$-band light curves from OGLE-II (data points),
plotted over 1.5 cycles, and model fits from \textsc{light2} 
(solid lines) for 10 SMC binaries;  note that
the vertical scale varies between panels. The (O$-$C) scatter about
the fit to each light curve is shown in the lower sub panel.}
 \label{lcplot1}
\end{figure*}

\begin{figure*}
\includegraphics[width=170mm]{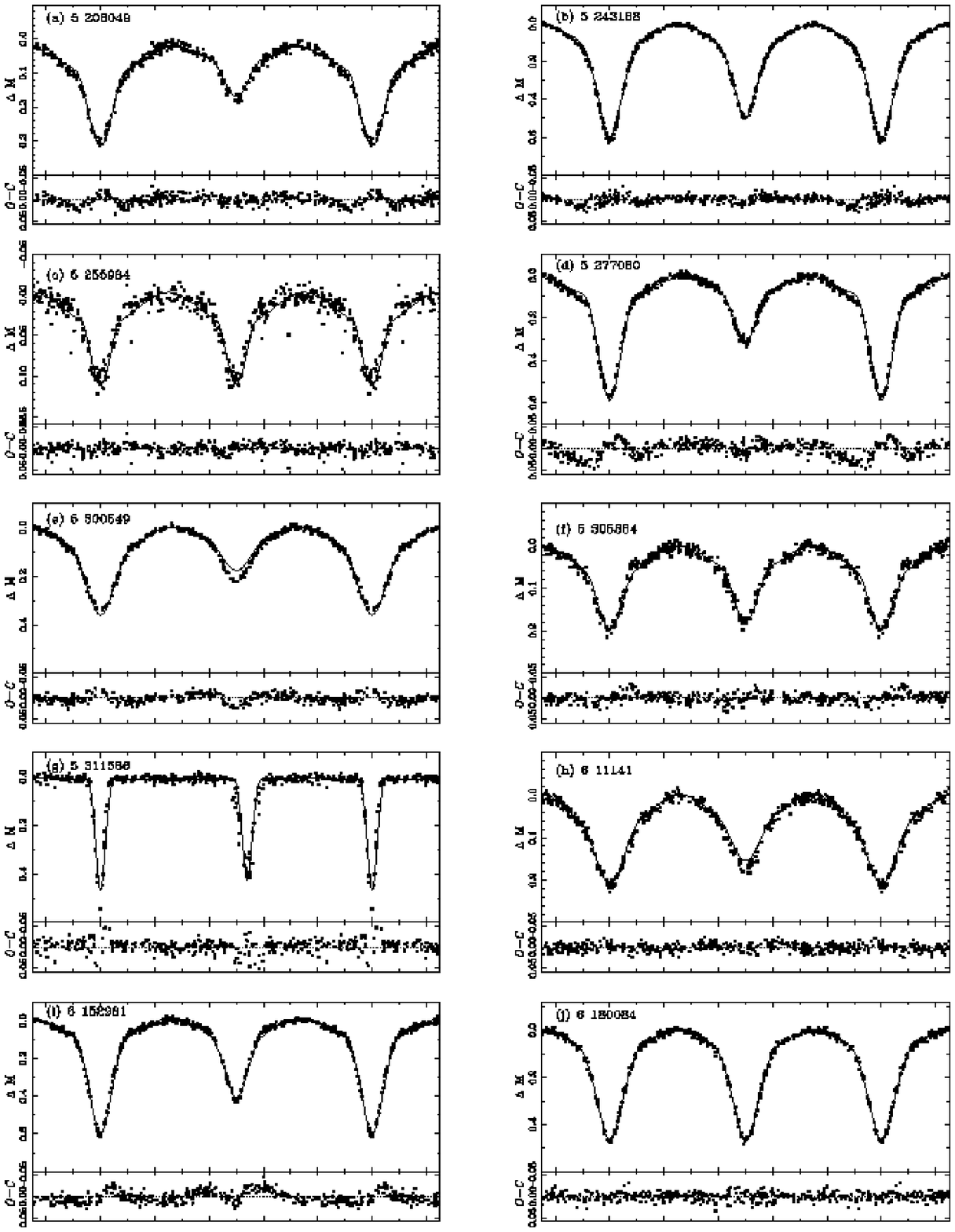} 
 \caption{$I$-band light curves from OGLE-II (data points),
plotted over 1.5 cycles, and model fits from \textsc{light2} 
(solid lines) for 10 SMC binaries;  note that
the vertical scale varies between panels. The (O$-$C) scatter about
the fit to each light curve is shown in the lower sub panel.}
 \label{lcplot2}
\end{figure*}

\begin{figure*}
\includegraphics[width=170mm]{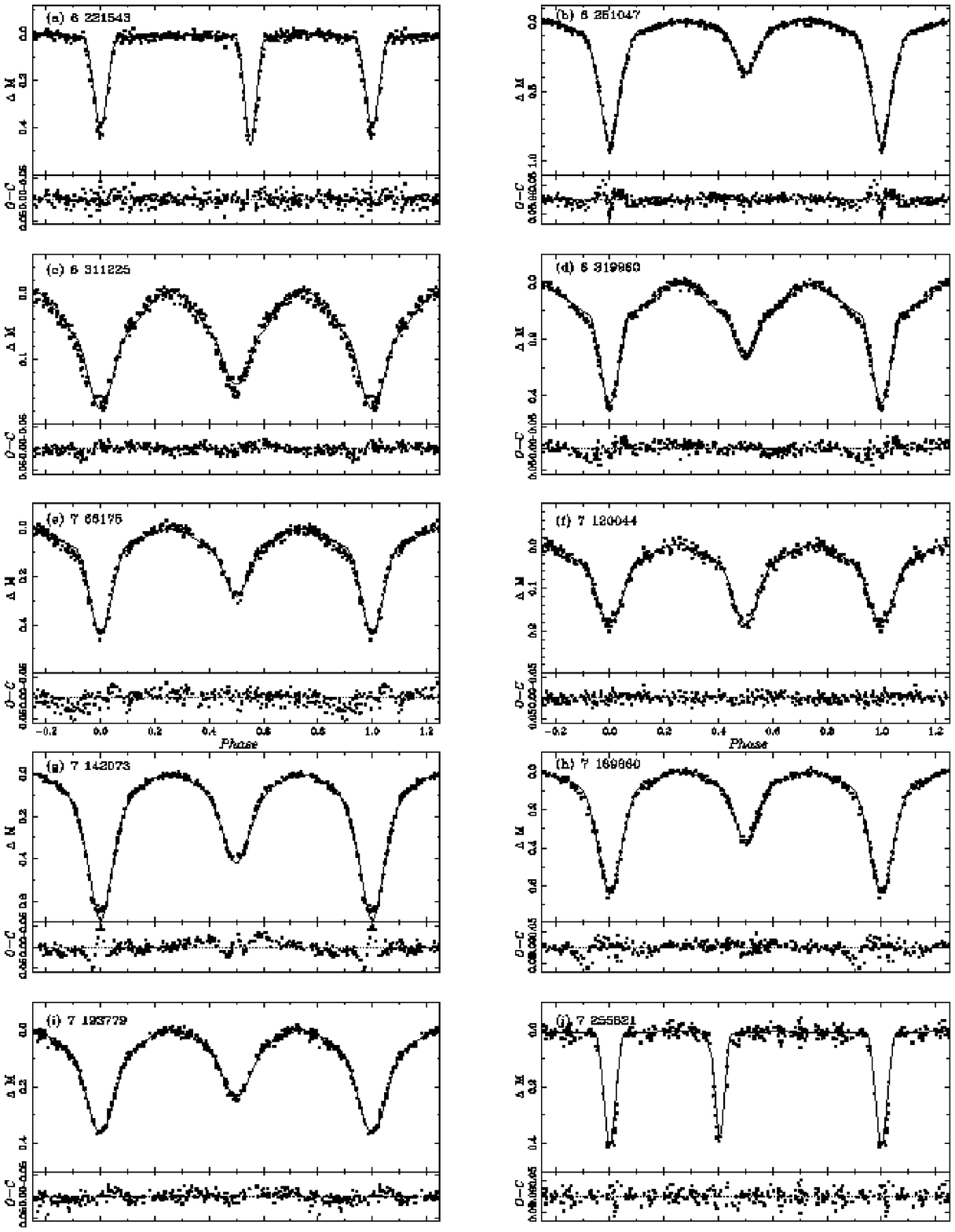} 
 \caption{$I$-band light curves from OGLE-II (data points),
plotted over 1.5 cycles, and model fits from \textsc{light2} 
(solid lines) for 10 SMC binaries;  note that
the vertical scale varies between panels. The (O$-$C) scatter about
the fit to each light curve is shown in the lower sub panel.}
 \label{lcplot3}
\end{figure*}

\begin{figure*}
\includegraphics[width=170mm]{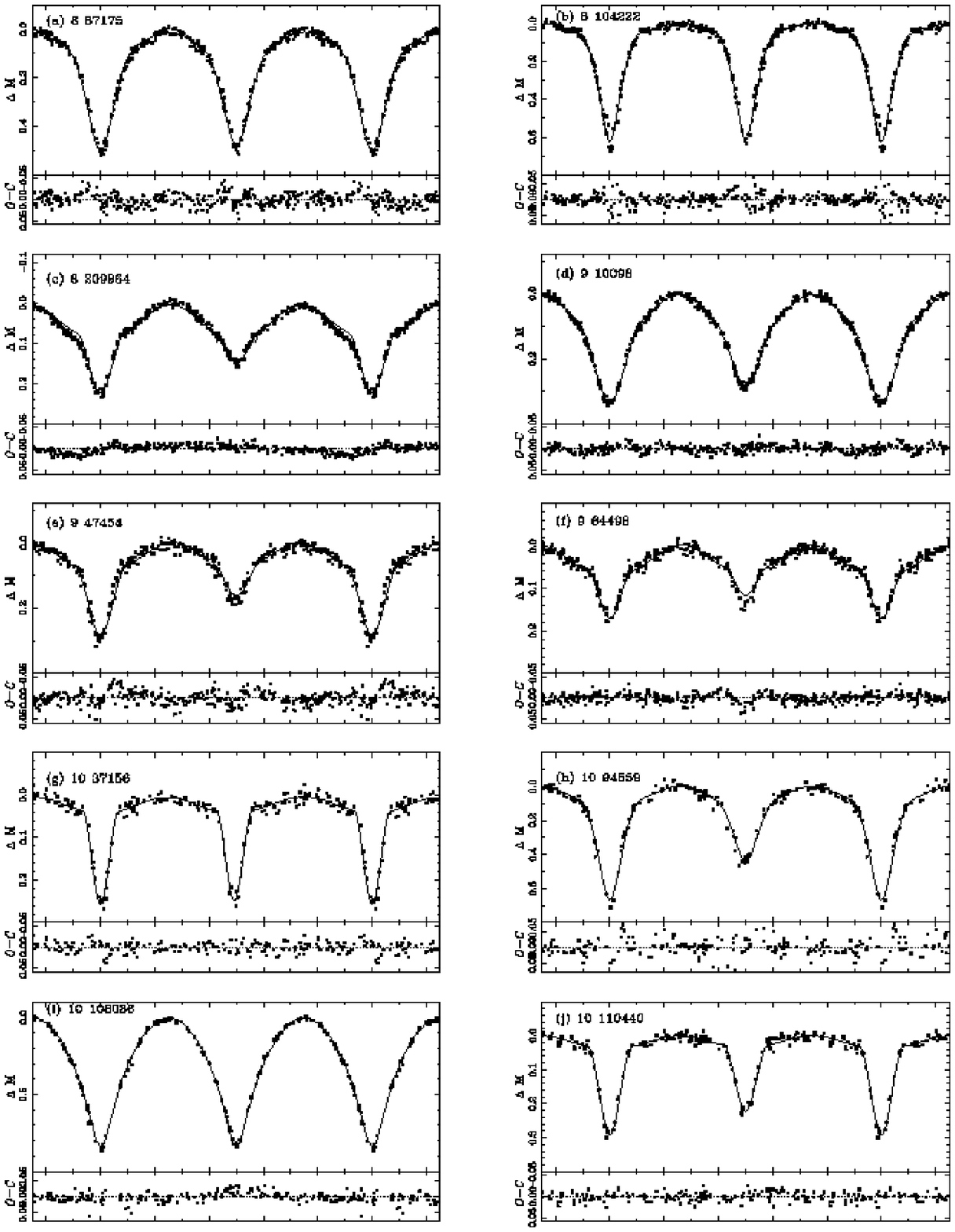} 
 \caption{$I$-band light curves from OGLE-II (data points),
plotted over 1.5 cycles, and model fits from \textsc{light2} 
(solid lines) for 10 SMC binaries;  note that
the vertical scale varies between panels. The (O$-$C) scatter about
the fit to each light curve is shown in the lower sub panel.}
 \label{lcplot4}
\end{figure*}

Table~\ref{lcsol} gives the adopted values for optimized parameters,
and the observed and model light curves are shown in
Figs~\ref{lcplot1}--\ref{lcplot4}, together with the corresponding
(O$-$C) plots. In 90\%\ of cases, the (O$-$C) curves show no
systematic phase-dependent deviations, demonstrating that satisfactory
fits have been obtained; the few exceptions are discussed
individually below.  We provide reduced chi-squared values,
$\chi^{2}_{\nu}$, in Table~\ref{lcsol}, as a quantitative measure of
the quality of each fit to each set of data.

\begin{table*}
\caption[]{Light-curve solutions.  
System types (column~2) are detached, semi-detached, or (in one case)
in contact.
Primary temperatures are based on
spectral types;  secondary temperatures characterize the $I$-band flux
ratio (Section~\ref{lcanal}).  Radii are given in units of the
semi-major axis of the relative orbit; for semi-detached systems these
are tied to the mass ratio.
}
\label{lcsol}
\begin{tabular}{rcllllllccc}
\hline
\multicolumn{1}{c}{OGLE-PSF}&
\multicolumn{1}{c}{System}&
\multicolumn{2}{c}{Spectral Type}&
\multicolumn{2}{c}{$T_{\rm eff}$ (K)}&
\multicolumn{2}{c}{Mean  radius}&
\multicolumn{1}{c}{Inclination}&
\multicolumn{1}{c}{Mass}&
$\chi^{2}_{\nu}$ \\
\multicolumn{1}{c}{identifier}&
\multicolumn{1}{c}{Type}&
\multicolumn{1}{c}{primary}&
\multicolumn{1}{c}{secondary}&
\multicolumn{1}{c}{primary}&
\multicolumn{1}{c}{secondary}&
\multicolumn{1}{c}{primary}&
\multicolumn{1}{c}{secondary}&
$(^{\circ})$&
\multicolumn{1}{c}{Ratio}& \\
\hline
%003851.98$-$733433.2&
 1~099121&$sd$&B0.5&B1--3&27800&$15820\pm165$&$0.248\pm0.006$&$0.332$&$85.4\pm0.3$&$1.704$&$2.54$ \\
%004633.14$-$732217.0&
 4~056804&$d$&B0&B0--2&30100&$24200\pm350$&$0.307\pm0.006$&$0.368\pm0.005$&$70.5\pm0.2$&$0.912$&$0.73$ \\
%004725.50$-$732716.7&
 4~103706&$d$&B0.5&B0--1&27800&$27050\pm430$&$0.345\pm0.007$&$0.261\pm0.007$&$75.8\pm0.4$&$1.768$&$1.29$ \\
%004700.19$-$731843.1&
 4~110409&$sd$&B1&B0.5-3&25500&$15390\pm160$&$0.177\pm0.007$&$0.341$&$76.8\pm0.3$&$1.544$&$2.80$ \\
%004753.24$-$731556.5&
 4~163552&$d$&B1&B0--3&25500&$25390\pm190$&$0.320\pm0.003$&$0.312\pm0.004$&$78.3\pm0.1$&$1.070$&$0.82$ \\
%004859.84$-$731328.8&
 5~026631&$sd$&B1&B1--3&25500&$17130\pm300$&$0.341\pm0.002$&$0.377$&$61.5\pm0.1$&$1.021$&$0.92$ \\
%004835.40$-$725256.5&
 5~060548&$sd$&B0&B0.5--2&30100&$17480\pm250$&$0.314\pm0.002$&$0.360$&$65.2\pm0.1$&$1.238$&$1.92$ \\
%004950.42$-$731931.6&
 5~095194&$d$&O9&O9&33800&$32600\pm100$&$0.252\pm0.002$&$0.298\pm0.002$&$65.3\pm0.1$&$0.870$&$0.39$ \\
%004943.08$-$725109.0&
 5~140701&$d$&B1.5&B2&23500&$15310\pm230$&$0.345\pm0.005$&$0.307\pm0.008$&$81.7\pm0.6$&$1.307$&$1.52$ \\
%005044.74$-$731739.9&
 5~180064&$d$&B1&B1&25500&$17110\pm1210$&$0.275\pm0.013$&$0.222\pm0.014$&$64.6\pm0.2$&$1.521$&$0.53$ \\
%005045.00$-$725844.4&
 5~208049&$sd$&B1&B1--5&25500&$14270\pm160$&$0.306\pm0.002$&$0.315$&$67.6\pm0.1$&$2.093$&$0.86$ \\
%005118.78$-$733015.8&
 5~243188&$sd$&O8.5&O9--B0&35300&$32110\pm50$&$0.319\pm0.003$&$0.345$&$79.9\pm0.1$&$1.466$&$1.01$ \\
%005129.62$-$732137.7&
 5~255984&$d$&B1&B1--2&25500&$25050\pm950$&$0.314\pm0.019$&$0.255\pm0.015$&$64.8\pm0.2$&$1.659$&$0.81$ \\
%005111.68$-$730520.3&
 5~277080&$sd$&B1&B2&25500&$15890\pm150$&$0.253\pm0.004$&$0.341$&$76.8\pm0.2$&$1.540$&$1.55$ \\
%005123.57$-$725224.1&
 5~300549&$sd$&B0&B0--2&30100&$16050\pm260$&$0.357\pm0.003$&$0.346$&$66.9\pm0.1$&$1.458$&$0.78$ \\
%005120.17$-$724942.9&
 5~305884&$d$&O9&O9&33800&$32500\pm240$&$0.340\pm0.005$&$0.286\pm0.013$&$65.8\pm0.6$&$1.086$&$1.63$ \\
%005134.85$-$724545.9&
 5~311566&$d$&B0&B0--3&30100&$28730\pm650$&$0.165\pm0.004$&$0.121\pm0.007$&$86.4\pm0.9$&$1.261$&$5.57$ \\
%005203.96$-$731849.3&
 6~011141&$d$&B0&B0--1&30100&$22740\pm460$&$0.350\pm0.002$&$0.348\pm0.003$&$61.3\pm0.2$&$1.068$&$0.64$ \\
%005241.88$-$724622.4&
 6~152981&$d$&B1&B1--3&25500&$20160\pm150$&$0.270\pm0.003$&$0.341$&$79.9\pm0.1$&$1.533$&$1.75$ \\
%005342.42$-$732319.9&
 6~180084&$d$&B0.5&B0--2&27800&$27930\pm350$&$0.346\pm0.005$&$0.337\pm0.006$&$77.0\pm0.1$&$1.169$&$0.79$ \\
%005340.40$-$725222.0&
 6~221543&$d$&B1&B1--2&25500&$26600\pm230$&$0.193\pm0.014$&$0.166\pm0.016$&$84.1\pm0.4$&$1.025$&$0.91$ \\
%005344.05$-$723124.0&
 6~251047&$sd$&B1.5&B3&23500&$13520\pm90$&$0.249\pm0.005$&$0.345$&$82.5\pm0.2$&$1.473$&$1.01$ \\
%005402.00$-$724221.6&
 6~311225&$sd$&B0&B0.5--2&30100&$23500\pm280$&$0.323\pm0.001$&$0.329$&$61.2\pm0.1$&$1.773$&$0.89$ \\
%005405.26$-$723426.0&
 6~319960&$sd$&B1&B1--2&25500&$16300\pm150$&$0.162\pm0.003$&$0.339$&$75.5\pm0.2$&$1.569$&$1.57$ \\
%005438.22$-$723206.2&
 7~066175&$sd$&O9.5&B0.5--1&32200&$25350\pm410$&$0.253\pm0.003$&$0.333$&$74.1\pm0.1$&$1.703$&$4.79$ \\
%005531.57$-$724307.8&
 7~120044&$d$&B1&B1&25500&$25560\pm410$&$0.325\pm0.005$&$0.318\pm0.007$&$65.1\pm0.1$&$1.004$&$0.81$ \\
%005554.44$-$722808.5&
 7~142073&$sd$&B0&B0.5--1.5&30100&$21010\pm280$&$0.394\pm0.003$&$0.319$&$81.2\pm0.1$&$2.000$&$2.52$ \\
%005637.30$-$724143.3&
 7~189660&$sd$&B1&B1--2&25500&$16910\pm160$&$0.307\pm0.004$&$0.343$&$78.1\pm0.1$&$1.498$&$1.52$ \\
%005621.80$-$723701.7&
 7~193779&$sd$&B1&B1--2&25500&$17160\pm220$&$0.377\pm0.003$&$0.321$&$69.5\pm0.1$&$1.966$&$0.94$ \\
%&
 7~255621&$d$&B1&B1--3&25500&$23370\pm890$&$0.177\pm0.009$&$0.125\pm0.020$&$85.4\pm1.5$&$1.293$&$1.64$ \\
%005830.98$-$723913.8&
 8~087175&$d$&B1&B1--2&25500&$24740\pm510$&$0.353\pm0.010$&$0.345\pm0.011$&$77.0\pm0.2$&$1.119$&$1.43$ \\
%005825.08$-$721909.8&
 8~104222&$d$&B1&B1--2&25500&$25260\pm790$&$0.315\pm0.021$&$0.312\pm0.021$&$84.1\pm0.3$&$1.078$&$2.03$ \\
%010016.05$-$721243.9&
 8~209964&$sd$&O8&O9&36300&$26460\pm270$&$0.255\pm0.005$&$0.355$&$64.4\pm0.4$&$1.300$&$1.38$ \\
%010052.90$-$724748.6&
 9~010098&$sd$&O9&B0&33800&$31840\pm90$&$0.357\pm0.001$&$0.355$&$67.5\pm0.1$&$1.304$&$0.74$ \\
%010052.04$-$720705.5&
 9~047454&$sd$&B1&B1--2&27800&$16890\pm300$&$0.264\pm0.003$&$0.352$&$66.9\pm0.2$&$1.362$&$1.94$ \\
%010117.26$-$724232.1&
 9~064498&$sd$&B1&B0.5--2&25500&$17060\pm360$&$0.301\pm0.009$&$0.285$&$64.6\pm0.4$&$3.092$&$0.80$ \\
%&
10~037156&$d$&O9.5&B0.5--2&32200&$31300\pm370$&$0.266\pm0.005$&$0.163\pm0.017$&$78.0\pm1.0$&$1.147$&$1.06$ \\
%&
10~094559&$sd$&B0&B1&30100&$23890\pm360$&$0.273\pm0.007$&$0.362$&$80.6\pm0.4$&$1.208$&$3.96$ \\
%&
10~108086&$c$&O9&O9.5-B0&33800&$29900\pm1000$&$0.466\pm0.002$&$0.438\pm0.002$&$82.8\pm0.1$&$1.183$&$1.88$ \\
%&
10~110440&$d$&B1&B2&25500&$21400\pm470$&$0.274\pm0.014$&$0.253\pm0.010$&$74.5\pm0.3$&$1.825$&$0.48$ \\
\hline
\end{tabular}
\end{table*}

\subsection{Notes on individual binaries}
\label{Notes}

In this section we comment on various aspects of individual solutions.
Targets are identified by entries in the Udalski et al.\ (1998)
catalogue and, in parentheses, the OGLE DIA catalogue 
(Zebrun {\em et al.}\ 2001).

Of the systems considered here, several have had light-curve solutions
attempted by Clausen et al. (2003; 5~140701), by Wyithe \& Wilson (2001,
2002; 4~103706, 4~163552, 7~66175, 7~255621, 8~87175), or by Graczyk
(2003; 4~103706, 4~163552, 5~311566, 6~221543, 8~104222).  Because
their solutions were unconstrained by spectroscopy, we do not make
detailed comparisons with, or comments on, their results, which must
be regarded as substantially less secure than ours.

\subsubsection{1~99121 (DIA 003851.98$-$733433.2)}
\label{X99121}

From 11 spectra in the wider phase ranges 0.15--0.35 and 0.65--0.85, a
single minimum was found in the disentangling solution. The
light-curve solution converged immediately to a semi-detached
configuration, with a photometric flux ratio of 1.2 confirmed by the relative
strengths of the spectral lines. The model fits the eclipse curves
well, but the observational data just before first contact of primary
eclipse are depressed relative to those just after fourth contact, and
this asymmetry is reflected in the value of $\chi_{\nu}^{2}$. Such
depressions in light curves occurring just before the onset of primary
eclipse are observed quite frequently in semi-detached (Algol-type)
binaries and are indicative of absorption by the gas stream from the
inner Lagrangian point of the Roche-lobe-filling secondary star falling 
towards the primary star. Just before primary eclipse, the gas stream is
seen in projection across the face of the primary component, and
causes the extra absorption.

\subsubsection{4~56804 (DIA 004633.14$-$732217.0)}

Spectra in the wider phase ranges yielded multiple $\chi^{2}$ minima
in the disentangling solutions, none of which gave physically sensible
results, but the four spectra closest to quadratures gave a mass ratio
(primary/secondary) $q<1.0$, and the resultant light-curve solution
provides an excellent fit to the data ($\chi^{2}_\nu=0.73$)
and a $B$-band flux ratio of unity, consistent with the spectra.

\subsubsection{4~103706 (DIA 004725.50$-$732716.7)}

A simple disentangling solution from spectra in the narrower phase
ranges (0.20--0.30, 0.70--0.80) coupled with a detached solution to
this well-defined light curve provides a flat (O$-$C) curve, rms
scatter of 0.013 mag, and $\chi^{2}_\nu=1.29$. The light-curve solution
predicts a $B$-band flux ratio of 1.8, consistent with the spectra. 

\subsubsection{4~110409 (DIA 004700.19$-$731843.1)}

There were no spectroscopic observations very close to phases 0.25 or 0.75,
but by using the complete spectra in the wider phase ranges a unique
solution was found, at a mass ratio of 1.544.  The light curve is
asymmetric, with the first maximum being 0.03 mag brighter than the
second, which limits the quality of the light-curve fit, as reflected
in the reduced chi-squared of 2.80. Nevertheless, the shapes of both
eclipses are well modeled, providing sensible stellar radii for this
semi-detached system. The implied photometric flux ratio of 0.6 agrees
very well with the spectroscopic value.

\subsubsection{4~163552 (DIA 004753.24$-$731556.5)}

An obviously double-lined system giving a simple disentangling
solution and a good detached solution to the well-defined light curve,
yielding an rms scatter of 0.011 mag and $\chi^{2}_\nu=0.82$. The
photometric flux ratio of 1.0 agrees with the spectra.

\subsubsection{5~26631 (DIA 004859.84$-$731328.8)}

Disentangling solutions of the 10 spectra in the narrower phase ranges
resulted in several $\chi^{2}$ minima, but only one (with a mass ratio
of unity) gave physically sensible results when combined with the
light-curve analysis.  Statistically indistinguishable solutions to
the light curve were obtained with a detached configuration (the
primary being the larger star), and with a semi-detached configuration
(with the secondary being larger). In the detached case, the primary
fills 97\% of its Roche lobe, the expected $B$-band flux ratio is 2.7,
and the resultant astrophysical parameters are inconsistent with
evolution models for single stars. In the semi-detached case, the
secondary fills its Roche lobe and the $B$-band flux ratio of 1.4 is
consistent with the spectra.  The s-d solution was therefore adopted,
giving a light-curve fit showing an rms scatter of 0.016 mag and
$\chi^{2}_\nu=0.92$.

\subsubsection{5~60548 (DIA 004835.40$-$725256.5)}

A simple disentangling solution was obtained from 22 spectra in the
wider phase ranges.  The light-curve solution converged to a
semi-detached configuration with a photometric flux ratio of 1.7,
consistent with the spectra.  Although the rms scatter of the
observations about the model fit is small, at 0.011 mag, the
$\chi^{2}_\nu$ value of 1.92 reflects the small asymmetry in the
observed curve just before first contact of primary eclipse.

\subsubsection{5~95194 (DIA 004950.42$-$731931.6)}

The two O-type stars are clearly seen in double-lined spectra,
with a spectroscopic flux ratio of 0.8. The disentangling solution was
obtained from 12 spectra in the wider phase ranges, mostly
covering first quadrature, and demonstrate that the secondary
component is the more massive star. The light-curve solution
($\chi^{2}_\nu=0.39$) was then determined with a secondary component
larger than the primary but a little cooler, as indicated by primary
eclipse being $\sim0.015$ mag deeper than secondary eclipse. The
resulting $B$-band flux ratio is consistent with that seen in the
spectra.

\subsubsection{5~140701 (DIA 004943.08$-$725109.0, HV~11284)}

A simple disentangling solution from 12 spectra in the wider phase
ranges led to a detached configuration from the light-curve solution,
and a photometric flux ratio of 2.3, consistent with the spectra. The
rms scatter of 0.011 and $\chi^{2}_\nu$ of 1.52 indicate a good solution.

\subsubsection{5~180064 (DIA 005044.74$-$731739.9)}

The 15 spectra covering both quadratures in the narrower phase ranges
provided a simple and well-defined disentangling solution, despite the
substantial flux ratio ($>2$). The light curve has a
low amplitude (0.08 mag), but the model fit is
excellent, with an rms scatter of 0.009 mag, and $\chi^{2}_\nu$ of 0.53. The
predicted $B$-band flux ratio of 2.7 is consistent with the spectra.

\subsubsection{5~208049 (DIA 005045.00$-$725844.4)}

All but one of the 16 spectra used in the first disentangling solution
lie in the phase range 0.65--0.85, and several minima were found. The
quoted result is from the spectra in the phase range 0.70--0.80, which
provided the best light-curve solution, yielding a semi-detached
configuration with a photometric flux ratio of 1.4 (consistent with
the spectra). The quality of the fit is reasonable, with rms scatter
of 0.016 mag, and $\chi^{2}_\nu=1.67$.

\subsubsection{5~243188 (DIA 005118.78$-$733015.8)}

The best disentangling solution was obtained with 18 spectra from the
wider phase ranges of 0.15--0.35 and 0.65--0.85, and the
large-amplitude light curve yielded alternative detached and
semi-detached solutions with different flux ratios.  The semi-detached
solution requires a flux ratio of 0.85, which is consistent with the
spectra (the secondary's spectral lines are slightly stronger than
those of the primary). The quality of the fit to the light curve is
excellent, with an rms scatter of 0.009 mag and $\chi^{2}_\nu=1.01$,
and a small dip just before first contact of primary eclipse.

\subsubsection{5~255984 (DIA 005129.62$-$732137.7)}

Spectra in the phase ranges 0.20--0.30 and 0.70--0.80 yielded 
a single sensible
disentangling solution, and the low-amplitude (0.12-mag) light curve
provides a good detached solution with a flat (O$-$C) curve, rms
scatter of 0.011 mag, and $\chi^{2}_\nu=0.81$. The model flux ratio
of 1.5 is consistent with the spectra.

\subsubsection{5~277080 (DIA 005111.68$-$730520.3)}

A simple disentangling solution from 13 spectra in the narrower phase
ranges, and a semi-detached configuration for the light-curve solution,
provide the best match to the flux ratio of unity seen in the
spectra. The (O$-$C) curve has some deviations around first contact of
primary eclipse, reflected in the rms scatter of 0.016 mag and
$\chi^{2}_\nu=1.55$.  Nonetheless, the overall fit appears to be
good.

\subsubsection{5~300549 (DIA 005123.57$-$725224.1)}

There are indications of emission lines at H$\gamma$, and for the
disentangling solution we used only the He$\;${\sc i} lines, with
phase ranges 0.20--0.30 and 0.70--0.80.  The light-curve required a
phase shift of $-0.0050$ before a good semi-detached configuration was
obtained, with a photometric flux ratio of 1.2, matching the line
spectra. The rms scatter is 0.010 mag, with $\chi^{2}_\nu=0.78$.

\subsubsection{5~305884 (DIA 005120.17$-$724942.9)}

The 17 spectra in the narrower phase ranges give one physically
sensible solution, and the light-curve analysis provides a reasonable fit
with rms scatter of 0.011 mag, and $\chi^{2}_\nu=1.63$. The predicted
$B$-band flux ratio of 1.5 is consistent with the spectra.

\subsubsection{5~311566 (6~67289, DIA 005134.85$-$724545.9)}
\label{X311566}

This system falls in the overlap of two OGLE fields, and so has two
identifiers.  In our initial analyses we solved the PSF light curves
for `5~311566' and `6~67289' both separately and combined, obtaining
closely similar results.

The DIA light curve is very similar in appearance to the PSF results,
clearly showing an eccentric orbit with deep eclipses, but
unfortunately the number of observations defining the shapes of both
eclipses is very limited (Fig.~4g), and this factor constrains the
quality of the overall solution. The light curve was solved as an
obvious detached system at a mass ratio of unity for the usual set of
parameters together with $e, \omega$. An orbital solution with these
$e, \omega$ values fixed provided the semi-amplitudes, and the
light-curve solution was slightly revised. The paucity of observations
through eclipses may contribute to the high rms of 0.036 mag and
$\chi^{2}_\nu=5.57$, by far the worst in the entire sample.

\subsubsection{6~11141 (DIA 005203.96$-$731849.3)}

A simple disentangling solution allowed a straightforward solution for
the light curve, which has a rather low amplitude of 0.22 mag. The
(O$-$C) curve is flat, with an rms scatter of 0.009 mag and
$\chi^{2}_\nu=0.64$. The photometric flux ratio of 1.7 is consistent
with the spectra.

\subsubsection{6~152981 (DIA 005241.88$-$724622.4)}

Only the first-quadrature phases for this binary were observed, but an
orbit solution was obtained straightforwardly.  The light curve has
deep eclipses, and a semi-detached configuration was required to
obtain the $B$-band flux ratio of 0.8 seen in the spectra. The rms
scatter about the model fit is 0.014 mag, whilst $\chi^{2}_\nu=1.75$
reflects apparent slight asymmetries in the eclipse curves. A slight
phase shift ($\sim\,0.002$) may improve the fit.

\subsubsection{6~180084 (DIA 005342.42$-$732319.9)}

Only 8 spectra in the wider phase ranges were available for this
system, but these gave a unique solution, indicating stars of similar
masses. The well-defined light curve with deep eclipses is accurately
modeled, giving an rms scatter of 0.014 mag and $\chi^{2}_\nu=0.79$,
and the photometric flux ratio of unity is confirmed by the spectra.

\subsubsection{6~221543 (DIA 005340.40$-$725222.0)}

For the second system in the sample with an obviously eccentric orbit,
the light curve was solved at an initial mass ratio of unity and
including $e, \omega$ as free parameters. The spectroscopic solution
was obtained with these $e, \omega$ values fixed, from 22 spectra. The
derived mass ratio of 1.025 clearly did not significantly change the
subsequent re-solution of the light curve, which yields an rms scatter
of 0.015 mag and $\chi^{2}_\nu=0.91$. The photometric flux ratio of 1.3 is
consistent with the spectra.

\subsubsection{6~251047 (DIA 005344.05$-$723124.0)}

A simple disentangling solution was secured, and the subsequent
light-curve solution proved to be straightforward, with a semi-detached
configuration providing a photometric flux ratio of 1.1. The DIA light
curve is free of the obvious contamination seen in the PSF results,
although the derived parameters are similar. A phase shift of 0.0020
was applied to the DIA light curve to improve the fit to the light
curve, which shows some anomalies in primary eclipse but a good fit
through secondary eclipse. The rms scatter is 0.022 mag, and
$\chi^{2}_\nu=1.01$.

\subsubsection{6~311225 (DIA 005402.00$-$724221.6)}

A total of 20 spectra cover both quadratures well, but more than one
solution was found. The adopted light-curve solution is for a
semi-detached configuration with a $B$-band flux ratio of 1.4,
consistent with the spectra.  The rms scatter of 0.009 mag and
$\chi^{2}_\nu=0.89$ indicate a good solution to this shallow light
curve, which nevertheless has a small asymmetry in the descending
branch of primary eclipse.

\subsubsection{6~319960 (DIA 005405.26$-$723426.0)}

Only the second-quadrature phases were sampled, by 15 spectra, but a
solution was obtained straighforwardly. The spectra indicate a flux
ratio of 0.4 and a semi-detached configuration. This result is
confirmed in the light-curve solution, which yields a reasonable fit,
with rms scatter of 0.012 mag and $\chi^{2}_\nu=1.57$. Again, there is
a small dip just before first contact of primary eclipse.

\subsubsection{7~66175 (DIA 005438.22$-$723206.2)}

A total of 20 spectra in the wider phase ranges, but mostly from
second quadrature, were used to establish the orbital solution of this
obviously double-lined system with a flux ratio less than unity. The
light-curve solution gives a good match to the overall shape of the
curve, but there is a significant depression just before first contact
of primary eclipse, which is reflected in $\chi^{2}_\nu=4.79$,
although the rms scatter is tolerable at 0.016 mag. The light-curve
solution for a semi-detached configuration correctly predicts the
spectroscopic flux ratio.

\subsubsection{7~120044 (DIA 005531.57$-$724307.8)}

An obvious double-lined system with well-sampled quadratures yielded a
simple disentangling solution from 12 spectra. The light curve is also
well modeled, by a model with a detached configuration, giving a flat
(O$-$C) curve, rms scatter of 0.009 mag and $\chi^{2}_\nu=0.81$, and
a flux ratio of unity, matching the spectroscopic value.

\subsubsection{7~142073 (DIA 005554.44$-$722808.5)}

Various competing $\chi^{2}$ minima in the disentangling solution from
19 spectra in the wider phase ranges were reduced to one physically
sensible solution with the use of the 6 spectra in the narrower phase
ranges.  The resulting mass ratio of 2.0 provides a solution to the
well-defined light curve with deep eclipses as a semi-detached
configuration and a flux ratio of 2.7, consistent with the
spectra. The model fit is generally good, with rms scatter of 0.016
mag, but a $\chi^{2}_\nu$ as large as 2.52 reflects the fact that the primary
eclipse in the model is too deep by 0.05 mag.

\subsubsection{7~189660 (DIA 005637.30$-$724143.3)}

A set of 10 spectra in the wider phase ranges yielded a sensible orbit
solution. A phase shift of $+0.0050$ was required to improve the model
fit to the light curve, which has deep eclipses. The overall rms
scatter is 0.016 mag, and $\chi^{2}_\nu=1.52$ reflects the fact that
there is, again, a small dip in the observed curve just before first contact
of primary eclipse in this semi-detached system. The photometric flux
ratio of 1.3 is consistent with the spectra.

\subsubsection{7~193779 (DIA 005621.80$-$723701.7)}

Both quadrature phases were well sampled, with 21 spectra, and the 
orbit solution was simple. The light curve is well defined, and the 
model fit to the data is very good, yielding a flat (O$-$C) curve,
an rms scatter of 0.012 mag and  $\chi^{2}_\nu=0.94$, and a flux ratio 
of 2.3, which is consistent with the spectra.

\subsubsection{7~255621 (8~30634)}

A second target that occurs in two OGLE fields, and the third system
in this sample with an obviously eccentric orbit, 7~255621 has a large
orbital eccentricity ($e=0.198$).  As with the other eccentric-orbit
systems in this sample, the number of observations defining the narrow
eclipses is very limited, and this factor handicaps the accuracy of
the solution.  We solved the PSF light curves for 7~255621 and for
8~30634 separately, confirming agreement (particularly in the determinations of
$e, \omega$).  We then combined the data into a single light curve with
the 7~255621 ephemeris, since there were no obvious zero-point
differences in the two datasets.  The combined light curve was solved
for $e, \omega$, and the results adopted in the disentangling solution
using 11 spectra in the wider phase ranges.  The light curve was
re-solved after the orbit solution to yield a good fit to the
available data, with a flat O$-$C curve, an rms scatter of 0.021 mag,
and $\chi^{2}_\nu=1.64$. The photometric flux ratio of 2.3 is
consistent with the spectra.  This system is not included in the DIA
database.

\subsubsection{8~87175 (DIA 005830.98$-$723913.8)}

Limited spectroscopic data gave 7 double-lined spectra in the narrower
phase ranges to provide an orbital solution. The well-defined light
curve, with deep eclipses, yielded a detached configuration, with a
reasonably flat (O$-$C) curve, an rms scatter of 0.016 mag, and
$\chi^{2}_\nu=1.43$, and a flux ratio of 1.1, in agreement with the
spectra.

\subsubsection{8~104222 (DIA 005825.08$-$721909.8)}

A total of 11 double-lined spectra in the wider phase ranges were
available for the orbit solution. The light curve, with deep eclipses
of nearly equal depth, has some inconsistent observations through both
eclipses. The best solution, with a reasonably flat (O$-$C) curve,
yields an rms scatter of 0.023 mag, and $\chi^{2}_\nu=2.03$, reflecting
the scatter in observations through the eclipses.

\subsubsection{8~209964 (DIA 010016.05$-$721243.9)}

The obviously double-lined spectra sampled both quadratures with a
total of 14 spectra in the wider phase ranges. The light-curve
solution indicates a semi-detached configuration and gives a flat
(O$-$C) curve, an rms scatter of 0.009 mag, and
$\chi^{2}_\nu=1.38$. The photometric flux ratio of 0.9 is confirmed by the
spectra, and again there is a small dip in the observed light curve
just before first contact of primary eclipse.

\subsubsection{9~10098 (DIA 010052.90$-$724748.6)}

The narrower phase ranges have 7 spectra, yielding an orbit solution
with a flux ratio greater than unity. The light-curve solution
requires a semi-detached configuration to match this flux ratio,
giving a flat (O$-$C) curve, an rms scatter of 0.009 mag and
$\chi^{2}_\nu=0.74$.

\subsubsection{9~47454 (DIA 010052.04$-$720705.5)}

Six spectra observed at both quadratures provide the orbital solution.
The light curve appears to be symmetric but the observations have
substantial scatter, which is reflected in the rms scatter of 0.016
mag, and $\chi^{2}_\nu=1.94$. The photometric flux ratio of 1.1,
confirmed by the spectra, results from a semi-detached configuration;
a detached solution was also found, but the implied photometric flux
ratio of 4.6 is clearly ruled out by the spectra.

\subsubsection{9~64498 (DIA 010117.26$-$724232.1)}

A total of 10 spectra in the wider phase ranges was needed to provide
an orbital solution. The shallow light curve has an amplitude of only
0.17 mag, but is symmetric. The light-curve solution provides a
reasonable fit with a mostly flat (O$-$C) curve, an rms scatter of
0.011 mag and $\chi^{2}_\nu=0.80$. The photometric flux ratio of 1.8
is consistent with the spectra.

\subsubsection{10~37156}

Fourteen weakly double-lined spectra in the wider phase ranges
were used to find the orbital solution. The spectra show some emission
in H$\gamma$, but using only the He$\;${\sc i} lines gives essentially
the same solution.  The light curve is quite well defined, but has
minimal coverage of the eclipse curves. Initial solutions indicated
that secondary eclipse was slightly displaced from phase 0.50, and
re-solutions including $e, \omega$ yielded a sensible small
eccentricity of 0.013. The final solution provides a flat (O$-$C)
curve, with an rms scatter of 0.013 mag and $\chi^{2}_\nu=1.06$, and a
flux ratio of 2.9 that is consistent with the spectra.

\subsubsection{10~94559}

Clearly double-lined spectra with a flux ratio of unity provided an
orbital solution from 8 observations in the narrower phase ranges.
Although the PSF light curve has sparsely defined eclipses, the
light-curve solution shows that a semi-detached configuration is
necessary to provide a photometric flux ratio in agreement with the
spectra. The rms scatter is large, at 0.029 mag, and
$\chi^{2}_\nu=3.96$ confirms the limited quality of this star's light
curve.

\subsubsection{10~108086}

Only 6 double-lined spectra, which indicate a flux ratio near unity,
were available in the wider phase ranges to provide an orbital
solution. The light curve requires a contact configuration, and was
solved by Rucinski's routines for contact binaries incorporated in
{\sc light2}, to yield a fill-out factor of $F=1.70$ (as defined by
Mochnacki 1984). A phase shift of $+0.0040$ was
required to improve the model fit, resulting in a flat (O$-$C) curve,
rms scatter of 0.016 mag, and $\chi^{2}_\nu=1.88$.  This deep-contact
binary, comprising two O stars with masses of 17 and 14M$_{\odot}$,
has an orbital period of only 0.88 days -- a remarkable system.

\subsubsection{10~110440}

The spectra show emission in the H lines, and the orbital solution
(using the spectra in the wider phase ranges) is based only on the
He$\;${\sc i}~$\lambda$4471 line. The light curve is symmetric and
well defined, and the solution yields a flat (O$-$C) curve with an rms
scatter of 0.011 mag and $\chi^{2}_\nu=0.47$ -- one of the smallest in
the sample. The photometric flux ratio of 1.5 is confirmed by the
spectra.

\subsubsection{Other systems}

Of the remaining $\sim$60 binaries similarly investigated, many failed
to provide meaningful orbital solutions due to the specific
combinations of observed orbital phases and $S/N$ values achieved in
the spectra, as noted in Section~\ref{DatAnal}. Some spectra were
significantly contaminated by emission, particularly in the Balmer
lines, presumably from nebular sources. At the spectral resolution
employed in these fibre-fed spectra, we were unable to remove these
nebular lines unambiguously, and hence orbital solutions could not be
obtained reliably.  Nonetheless, there are evidently several
intrinsically interesting binary systems amongst this set, notably
6~89617 and 6~194856, both of which are composed of two O-type stars
with orbital periods of, respectively, 0.87 day and 0.65 day. Clearly,
such rare objects deserve to be studied more intensively in order to
establish their properties and to test models for the formation of
close binaries of high mass.

\section{Astrophysical Parameters and Evolutionary Characteristics}
\label{Params}

Astrophysical parameters can be calculated directly from the orbital
and the light-curve solutions.  Results are presented in
Table~\ref{detmrgtl} for the components in detached systems, and in
Table~\ref{sdmrgtl} for the semi-detached and contact systems. The
1-$\sigma$ uncertainties on the derived masses, radii, surface
gravities, temperatures and luminosities have been calculated by the
standard procedures of propagation of uncertainties (errors) from the
data given in Table~\ref{circorb}, Table~\ref{eccorb}, and
Table~\ref{lcsol} (cf., e.g., Hilditch 2001). For all systems, an
uncertainty of $\pm$1000\,K on the mean effective temperature of the
primary componentwas adopted, based on the discussion of the
spectral-type--effective-temperature calibration given in H$^3$03.

\subsection{Detached systems}
\label{DetSys}
\begin{table*}
\caption[]{Astrophysical parameters for detached systems}
\label{detmrgtl}
\begin{tabular}{rrrccccccccc}
\hline
\multicolumn{1}{c}{OGLE-PSF}&
\multicolumn{1}{c}{Mass}&
\multicolumn{1}{c}{Radius}&
\multicolumn{1}{c}{$\log\,g$}&
\multicolumn{1}{c}{$\log\,T_{\rm eff}$}&
\multicolumn{1}{c}{$\log\,L$}&
\multicolumn{1}{r}{$\delta\,M$}\\
\multicolumn{1}{c}{identifier}&
\multicolumn{1}{c}{(M$_{\odot}$)}&
\multicolumn{1}{c}{(R$_{\odot}$)}&
\multicolumn{1}{c}{(dex cgs)}&
\multicolumn{1}{c}{(dex K)}&
\multicolumn{1}{c}{(dex L$_{\odot}$)}&
\multicolumn{1}{c}{(M$_{\odot}$)}\\
\hline
%004633.14$-$732217.0&
 4~056804&$13.0\pm0.3$&$4.1\pm0.1$&$4.32\pm0.02$&$4.479\pm0.014$&$4.128\pm0.061$&+0.2 \\
$ $&$14.3\pm0.5$&$4.9\pm0.1$&$4.21\pm0.02$&$4.384\pm0.006$&$3.908\pm0.029$&-4.1 \\
%004725.50$-$732716.7&
 4~103706&$17.5\pm0.6$&$5.4\pm0.1$&$4.22\pm0.03$&$4.444\pm0.016$&$4.221\pm0.066$&-5.5 \\
$ $&$9.9\pm0.5$&$4.0\pm0.1$&$4.22\pm0.04$&$4.432\pm0.007$&$3.929\pm0.038$&+1.0 \\
%004753.24$-$731556.5&
 4~163552&$13.3\pm1.0$&$5.3\pm0.1$&$4.11\pm0.04$&$4.407\pm0.017$&$4.062\pm0.072$&-2.7 \\
$ $&$12.4\pm1.1$&$5.2\pm0.2$&$4.10\pm0.05$&$4.405\pm0.003$&$4.033\pm0.028$&-1.9 \\
%004859.84$-$731328.8&
 5~095194&$20.3\pm4.5$&$8.1\pm0.6$&$3.93\pm0.11$&$4.529\pm0.013$&$4.914\pm0.079$&+1.0 \\
$ $&$23.3\pm5.0$&$9.5\pm0.7$&$3.85\pm0.11$&$4.513\pm0.001$&$4.997\pm0.060$&-0.8 \\
%004943.08$-$725109.0&
 5~140701&$6.9\pm0.7$&$7.9\pm0.3$&$3.49\pm0.06$&$4.371\pm0.018$&$4.263\pm0.081$&+4.3 \\
$ $&$5.3\pm0.7$&$7.0\pm0.3$&$3.47\pm0.07$&$4.185\pm0.007$&$3.418\pm0.046$&+0.4 \\
%005044.74$-$731739.9&
 5~180064&$10.7\pm0.4$&$5.6\pm0.3$&$3.97\pm0.04$&$4.407\pm0.017$&$4.105\pm0.080$&+0.1 \\
$ $&$7.0\pm0.3$&$4.5\pm0.3$&$3.98\pm0.06$&$4.233\pm0.031$&$3.223\pm0.136$&-1.5 \\
%005129.62$-$732137.7&
 5~255984&$11.6\pm2.0$&$4.2\pm0.4$&$4.27\pm0.12$&$4.407\pm0.017$&$3.850\pm0.112$&-1.6 \\
$ $&$7.0\pm2.7$&$3.4\pm0.3$&$4.23\pm0.19$&$4.399\pm0.016$&$3.639\pm0.110$&+2.3 \\
%005120.17$-$724942.9&
 5~305884&$17.6\pm1.3$&$7.8\pm0.3$&$3.90\pm0.04$&$4.529\pm0.013$&$4.881\pm0.059$&+3.0 \\
$ $&$16.2\pm2.0$&$6.5\pm0.4$&$4.02\pm0.07$&$4.512\pm0.003$&$4.662\pm0.050$&+1.6 \\
%005134.85$-$724545.9&
 5~311566&$12.8\pm0.6$&$4.4\pm0.1$&$4.26\pm0.03$&$4.479\pm0.014$&$4.183\pm0.063$&+0.5 \\
$ $&$10.2\pm0.8$&$3.2\pm0.2$&$4.43\pm0.06$&$4.458\pm0.010$&$3.832\pm0.065$&+0.9 \\
%005203.96$-$731849.3&
 6~011141&$15.1\pm0.3$&$5.1\pm0.1$&$4.21\pm0.01$&$4.479\pm0.014$&$4.307\pm0.058$&-1.3 \\
$ $&$14.1\pm0.4$&$5.0\pm0.1$&$4.19\pm0.02$&$4.357\pm0.009$&$3.813\pm0.037$&-5.3 \\
%005342.42$-$732319.9&
 6~180084&$16.2\pm1.3$&$5.9\pm0.2$&$4.11\pm0.05$&$4.444\pm0.016$&$4.297\pm0.073$&-3.5 \\
$ $&$13.8\pm2.3$&$5.7\pm0.2$&$4.07\pm0.08$&$4.446\pm0.005$&$4.283\pm0.044$&-1.4 \\
%005340.40$-$725222.0&
 6~221543&$11.9\pm1.8$&$5.3\pm0.5$&$4.07\pm0.10$&$4.407\pm0.017$&$4.056\pm0.104$&-1.3 \\
$ $&$11.6\pm2.5$&$4.5\pm0.5$&$4.19\pm0.14$&$4.425\pm0.004$&$3.997\pm0.100$&-0.7 \\
%005531.57$-$724307.8&
 7~120044&$12.5\pm0.5$&$4.8\pm0.1$&$4.18\pm0.02$&$4.407\pm0.017$&$3.971\pm0.070$&-2.0 \\
$ $&$12.4\pm0.6$&$4.7\pm0.1$&$4.19\pm0.03$&$4.408\pm0.007$&$3.956\pm0.035$&-2.2 \\
%&
 7~255621&$9.3\pm0.8$&$5.0\pm0.3$&$4.00\pm0.06$&$4.407\pm0.017$&$4.016\pm0.084$&+1.3 \\
$ $&$7.2\pm0.6$&$3.5\pm0.6$&$4.20\pm0.14$&$4.369\pm0.017$&$3.558\pm0.153$&+1.0 \\
%005830.98$-$723913.8&
 8~087175&$12.0\pm1.0$&$4.5\pm0.2$&$4.21\pm0.06$&$4.407\pm0.017$&$3.915\pm0.080$&-1.5 \\
$ $&$10.7\pm1.6$&$4.4\pm0.2$&$4.18\pm0.08$&$4.393\pm0.009$&$3.843\pm0.056$&-0.5 \\
%005825.08$-$721909.8&
 8~104222&$13.1\pm0.9$&$5.2\pm0.4$&$4.11\pm0.07$&$4.407\pm0.017$&$4.052\pm0.093$&-2.5 \\
$ $&$12.1\pm1.3$&$5.2\pm0.4$&$4.09\pm0.08$&$4.402\pm0.014$&$4.026\pm0.084$&-1.5 \\
%&
10~037156&$19.5\pm0.4$&$7.2\pm0.2$&$4.02\pm0.02$&$4.508\pm0.013$&$4.730\pm0.057$&-1.0 \\
$ $&$17.0\pm0.5$&$4.4\pm0.5$&$4.38\pm0.09$&$4.496\pm0.005$&$4.257\pm0.093$&-2.7 \\
%&
10~110440&$10.8\pm2.1$&$4.0\pm0.4$&$4.27\pm0.12$&$4.407\pm0.017$&$3.810\pm0.110$&-0.8 \\
$ $&$5.9\pm2.3$&$3.7\pm0.3$&$4.08\pm0.19$&$4.330\pm0.011$&$3.436\pm0.089$&+1.0 \\
\hline
\end{tabular}
\end{table*}

\begin{figure*}
\includegraphics[width=150mm]{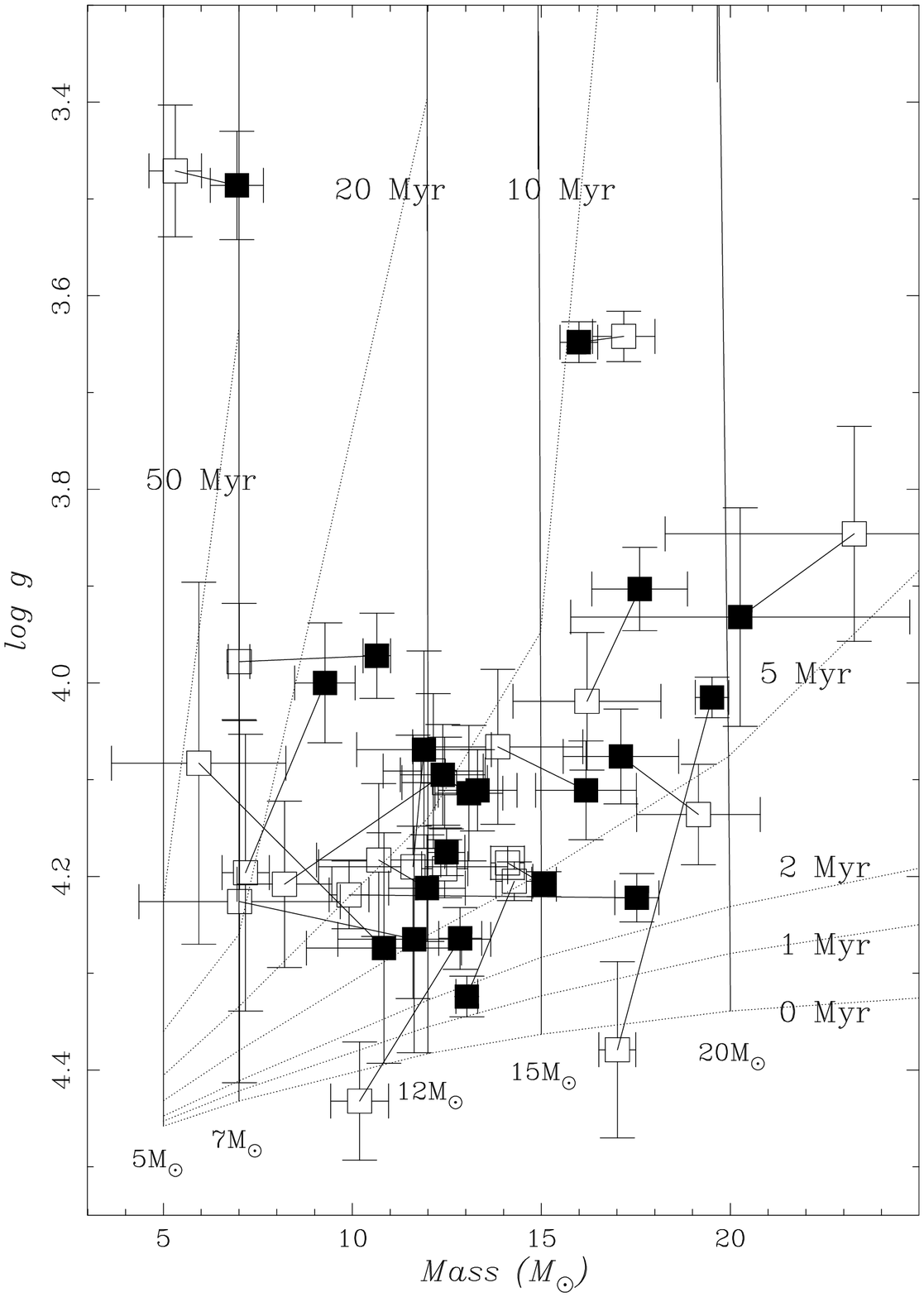} 
 \caption{The primary and secondary 
 components (filled and open symbols) of  21 SMC detached binary systems plotted with
 1-$\sigma$ uncertainties in the mass--surface-gravity plane (18 from
 this paper, 3 from H$^3$03). Lines join the primary and
 secondary of each binary. Also plotted are evolutionary tracks
 (solid lines) for single stars of masses 5, 7, 12, 15, and
 20M$_{\odot}$ and isochrones (dotted lines) for ages of 0, 1, 2,
 5, 10, and 20Myr, taken from the models of Girardi {\em et al.}\ (2000) for
 $Z=0.004$, appropriate for the SMC.}  
\label{detmlgg}
\end{figure*}

In any projection of the Hertzsprung-Russell diagram, the coeval
components of a pre-mass-exchange binary must lie on an isochrone.
This expectation is borne out for our sample of SMC detached systems,
as illustrated in Fig.~\ref{detmlgg}, where we compare our results for
detached systems with the stellar-evolution sequences given by Girardi
{\em et al.}\ (2000), choosing models for $Z=0.004$ as appropriate to
SMC metallicity, and the mass--surface-gravity plane for the
comparison.  The average mass ratio for the 21 binaries is 1.2, so
that differential evolution due to mass differences is not generally
expected to be important.  There are a few exceptions, notably
4~103706 (18+10M$_{\odot}$), 5~255984 (12+7M$_{\odot}$), and 10~110440
(11+6M$_{\odot}$), for which the locations of the components are
nonetheless consistent with isochrones to within the 1-$\sigma$
uncertainties.

\begin{figure*}
\includegraphics[width=150mm]{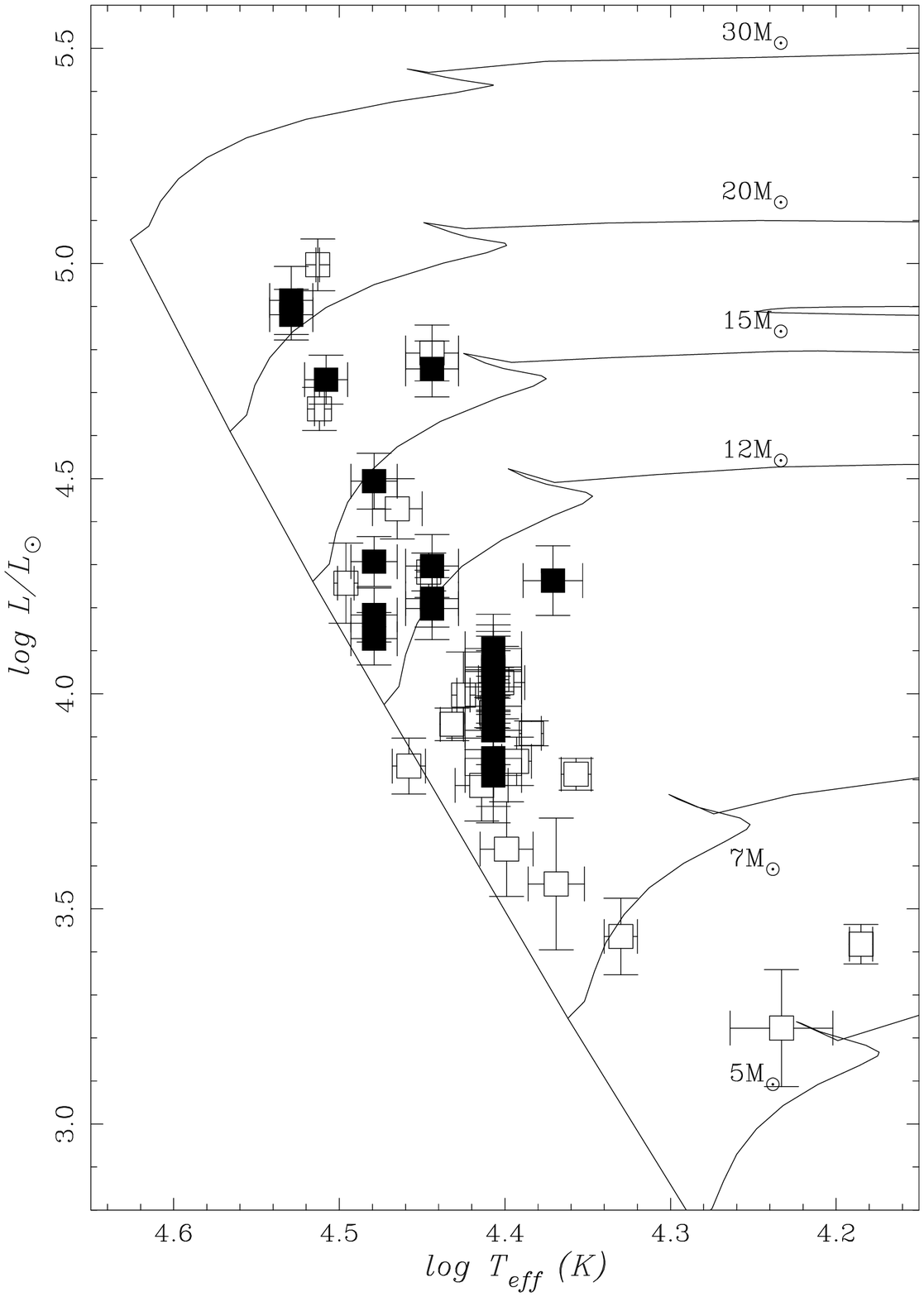} 
 \caption{The primary and secondary 
 components (filled and open symbols) of  21 SMC detached binary systems plotted with
 1-$\sigma$ uncertainties in the $\log\,T_{\rm
 eff}$--$\log\,L/$L$_{\odot}$ plane (HR diagram; 18 from this paper, 3
 from H$^3$03). Also plotted are the zero-age main sequence and
 evolutionary tracks for single stars of masses
 5, 7, 12, 15, 20, 30M$_{\odot}$, taken from the models of Girardi {\em et
 al.}\ (2000) for $Z=0.004$, appropriate for the SMC.}
\label{detlgtlgl}
\end{figure*}

A second, and more stringent, observational test of the models is to
compare the empirical, dynamically-determined masses with those
inferred from the evolutionary tracks in the HR diagram
(Fig.~\ref{detlgtlgl}). The differences between the evolutionary mass
and the observed mass for each of the 36 stars in detached systems are
listed in Table~\ref{detmrgtl}, in the sense evolutionary mass minus
observed mass. The average difference is
$\Delta\,M=-0.8\pm2.1\,M_{\odot}$ (s.d.), and when combined with the 6
stars in detached systems from H$^3$03, $\Delta\,M=-0.9\pm$2.2M$_{\odot}$
(s.d.). Considering that the typical uncertainties on the observed
masses are $\sim\pm1$--2M$_{\odot}$, this agreement is excellent, and
confirms that where discrepancies occur between `evolutionary' and 
`spectroscopic' masses (e.g., Herrero et al. 1992), it is the spectroscopic
results that are more questionable.

\begin{figure*}
\includegraphics[width=150mm]{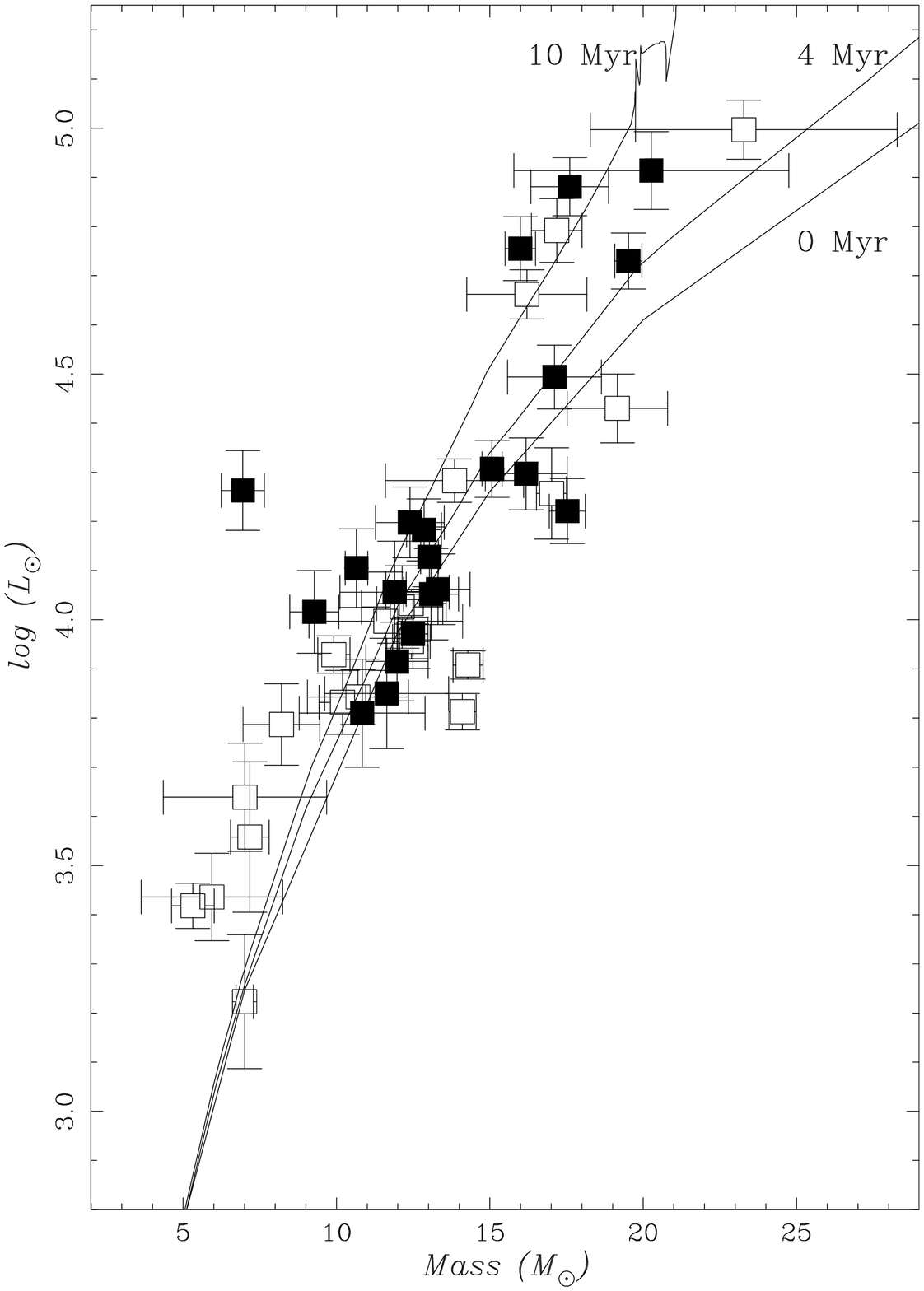} 
 \caption{The primary and secondary 
 components (filled and open symbols) of  21 SMC detached binary systems plotted with
 1-$\sigma$ uncertainties in the mass--$\log\,L/$L$_{\odot}$ plane (18
 from this paper, 3 from H$^3$03). Also plotted are  isochrones
 (solid lines) for ages of $0,4,10\,$Myr from the models of Girardi {\em et
 al.}\ (2000) for $Z=0.004$, appropriate for the SMC.}  
\label{detmlgl}
\end{figure*}

A final, straightforward comparison can be made by examining isochrones
in the mass--luminosity plane (Fig.~\ref{detmlgl});  all stars should
lie above the ZAMS.  In practice, most of the observed components
are distributed sensibly within their 1-$\sigma$ uncertainties of the
expectations from the models.  The obvious exceptions are the secondary
components of 4~56804, 6~11141, and 10~37156, and the primary component 
of 4~103706.

The overall conclusion from these comparisons is that agreement is
remarkably good between observations and the Girardi {\em et al.} models,
with only 4 exceptions in a sample of 44 stars.

\subsection{Semi-detached and contact systems}
\label{SDSys}
\begin{table*}
\caption[]{Astrophysical parameters for semi-detached and contact systems}
\label{sdmrgtl}
\begin{tabular}{rrrccccccccc}
\hline
\multicolumn{1}{c}{OGLE-PSF}&
\multicolumn{1}{c}{Mass}&
\multicolumn{1}{c}{Radius}&
\multicolumn{1}{c}{$\log\,g$}&
\multicolumn{1}{c}{$\log\,T_{\rm eff}$}&
\multicolumn{1}{c}{$\log\,L$}&
\multicolumn{1}{c}{$\delta\,M$}\\
\multicolumn{1}{c}{identifier}&
\multicolumn{1}{c}{(M$_{\odot}$)}&
\multicolumn{1}{c}{(R$_{\odot}$)}&
\multicolumn{1}{c}{(dex cgs)}&
\multicolumn{1}{c}{(dex K)}&
\multicolumn{1}{c}{(dex L$_{\odot}$)}&
\multicolumn{1}{c}{(M$_{\odot}$)}\\
\hline
%003851.98$-$733433.2&
 1~099121&$11.3\pm0.8$&$5.0\pm0.2$&$4.10\pm0.05$&$4.444\pm0.015$&$4.155\pm0.071$&+0.5 \\
$ $&$6.6\pm0.9$&$6.7\pm0.2$&$3.61\pm0.06$&$4.199\pm0.005$&$3.430\pm0.036$& \\
%004700.19$-$731843.1&
 4~110409&$13.7\pm0.8$&$4.3\pm0.2$&$4.30\pm0.05$&$4.407\pm0.017$&$3.886\pm0.079$&-5.6 \\
$ $&$8.9\pm1.1$&$8.4\pm0.3$&$3.54\pm0.06$&$4.187\pm0.045$&$3.580\pm0.035$& \\
%004859.84$-$731328.8&
 5~026631&$11.5\pm0.6$&$5.1\pm0.1$&$4.08\pm0.03$&$4.407\pm0.017$&$4.031\pm0.071$&-0.9 \\
$ $&$11.3\pm0.8$&$5.7\pm0.1$&$3.99\pm0.04$&$4.234\pm0.008$&$3.425\pm0.035$& \\
%004835.40$-$725256.5&
 5~060548&$10.8\pm0.4$&$8.4\pm0.1$&$3.62\pm0.02$&$4.479\pm0.014$&$4.748\pm0.059$&+6.7 \\
$ $&$8.7\pm0.4$&$9.6\pm0.2$&$3.41\pm0.03$&$4.243\pm0.006$&$3.923\pm0.028$& \\
%005045.00$-$725844.4&
 5~208049&$10.0\pm0.2$&$6.6\pm0.1$&$3.80\pm0.01$&$4.407\pm0.017$&$4.253\pm0.069$&+1.5 \\
$ $&$4.8\pm0.2$&$6.8\pm0.1$&$3.45\pm0.02$&$4.154\pm0.005$&$3.271\pm0.022$& \\
%005118.78$-$733015.8&
 5~243188&$27.3\pm1.5$&$7.3\pm0.2$&$4.15\pm0.03$&$4.550\pm0.012$&$4.912\pm0.054$&-4.7 \\
$ $&$18.7\pm1.9$&$7.9\pm0.2$&$3.91\pm0.05$&$4.507\pm0.001$&$4.807\pm0.022$& \\
%005111.68$-$730520.3&
 5~277080&$17.4\pm0.9$&$5.0\pm0.1$&$4.27\pm0.03$&$4.407\pm0.017$&$4.019\pm0.072$&-6.9 \\
$ $&$11.3\pm1.0$&$6.8\pm0.2$&$3.82\pm0.05$&$4.201\pm0.004$&$3.458\pm0.028$& \\
%005123.57$-$725224.1&
 5~300549&$25.4\pm1.0$&$6.4\pm0.1$&$4.24\pm0.02$&$4.479\pm0.014$&$4.506\pm0.060$&-10.4 \\
$ $&$17.4\pm1.1$&$6.1\pm0.1$&$4.10\pm0.03$&$4.205\pm0.007$&$3.385\pm0.032$& \\
%005241.88$-$724622.4&
 6~152981&$12.5\pm0.4$&$4.9\pm0.1$&$4.15\pm0.02$&$4.407\pm0.017$&$4.000\pm0.070$&-2.0 \\
$ $&$8.2\pm0.3$&$6.3\pm0.1$&$3.76\pm0.02$&$4.304\pm0.003$&$3.797\pm0.018$& \\
%005344.05$-$723124.0&
 6~251047&$8.1\pm0.2$&$4.6\pm0.1$&$4.02\pm0.02$&$4.371\pm0.018$&$3.799\pm0.077$&+0.9 \\
$ $&$5.5\pm0.2$&$6.4\pm0.1$&$3.57\pm0.02$&$4.131\pm0.003$&$3.123\pm0.020$& \\
%005402.00$-$724221.6&
 6~311225&$21.2\pm0.4$&$6.6\pm0.1$&$4.13\pm0.01$&$4.479\pm0.014$&$4.533\pm0.058$&-6.2 \\
$ $&$11.9\pm0.4$&$6.7\pm0.1$&$3.86\pm0.02$&$4.371\pm0.005$&$4.121\pm0.022$& \\
%005405.26$-$723426.0&
 6~319960&$10.6\pm0.8$&$4.5\pm0.2$&$4.16\pm0.05$&$4.407\pm0.017$&$3.913\pm0.077$&-2.4 \\
$ $&$6.7\pm1.0$&$9.4\pm0.3$&$3.32\pm0.07$&$4.212\pm0.004$&$3.779\pm0.035$& \\
%005438.22$-$723206.2&
 7~066175&$19.6\pm1.8$&$7.9\pm0.3$&$3.94\pm0.05$&$4.508\pm0.013$&$4.811\pm0.064$&-0.5 \\
$ $&$11.5\pm1.9$&$10.4\pm0.4$&$3.47\pm0.08$&$4.404\pm0.007$&$4.634\pm0.044$& \\
%005554.44$-$722808.5&
 7~142073&$12.6\pm1.2$&$9.6\pm0.3$&$3.58\pm0.05$&$4.479\pm0.014$&$4.862\pm0.065$&+6.2 \\
$ $&$6.3\pm0.9$&$7.8\pm0.3$&$3.46\pm0.07$&$4.322\pm0.006$&$4.054\pm0.039$& \\
%005637.30$-$724143.3&
 7~189660&$15.3\pm1.2$&$5.3\pm0.2$&$4.17\pm0.04$&$4.407\pm0.017$&$4.068\pm0.073$&-4.9 \\
$ $&$10.2\pm1.0$&$6.0\pm0.2$&$3.89\pm0.05$&$4.228\pm0.004$&$3.451\pm0.031$& \\
%005621.80$-$723701.7&
 7~193779&$11.6\pm1.0$&$5.8\pm0.2$&$3.98\pm0.05$&$4.407\pm0.017$&$4.138\pm0.075$&-0.6 \\
$ $&$5.9\pm0.9$&$4.9\pm0.2$&$3.82\pm0.07$&$4.235\pm0.006$&$3.310\pm0.039$& \\
%010016.05$-$721243.9&
 8~209964&$18.8\pm0.9$&$7.7\pm0.2$&$3.94\pm0.03$&$4.560\pm0.012$&$4.995\pm0.053$&+5.8 \\
$ $&$14.5\pm1.0$&$10.7\pm0.2$&$3.54\pm0.03$&$4.423\pm0.004$&$4.734\pm0.026$& \\
%010052.90$-$724748.6&
 9~010098&$17.8\pm1.8$&$5.1\pm0.2$&$4.28\pm0.06$&$4.529\pm0.013$&$4.514\pm0.063$&0.0 \\
$ $&$13.7\pm2.3$&$5.1\pm0.2$&$4.17\pm0.08$&$4.503\pm0.001$&$4.406\pm0.037$& \\
%010052.04$-$720705.5&
 9~047454&$12.6\pm2.3$&$4.2\pm0.3$&$4.29\pm0.10$&$4.444\pm0.016$&$4.008\pm0.083$&-1.2 \\
$ $&$9.2\pm2.0$&$5.6\pm0.3$&$3.91\pm0.11$&$4.228\pm0.008$&$3.390\pm0.062$& \\
%010117.26$-$724232.1&
 9~064498&$8.4\pm0.7$&$5.4\pm0.2$&$3.90\pm0.05$&$4.407\pm0.017$&$4.077\pm0.078$&+2.3 \\
$ $&$2.7\pm0.5$&$5.1\pm0.2$&$3.46\pm0.09$&$4.232\pm0.009$&$3.329\pm0.054$& \\
%&
10~094559&$12.0\pm1.0$&$4.7\pm0.2$&$4.18\pm0.05$&$4.479\pm0.014$&$4.240\pm0.070$&+1.4 \\
$ $&$10.0\pm1.4$&$6.2\pm0.3$&$3.85\pm0.07$&$4.378\pm0.065$&$4.083\pm0.044$& \\
%&
10~108086&$16.9\pm1.2$&$5.7\pm0.2$&$4.16\pm0.04$&$4.529\pm0.013$&$4.608\pm0.058$& \\
$ $&$14.3\pm1.7$&$5.3\pm0.2$&$4.14\pm0.06$&$4.476\pm0.015$&$4.341\pm0.064$& \\
\hline
\end{tabular}
\end{table*}

\begin{figure*}
\includegraphics[width=150mm]{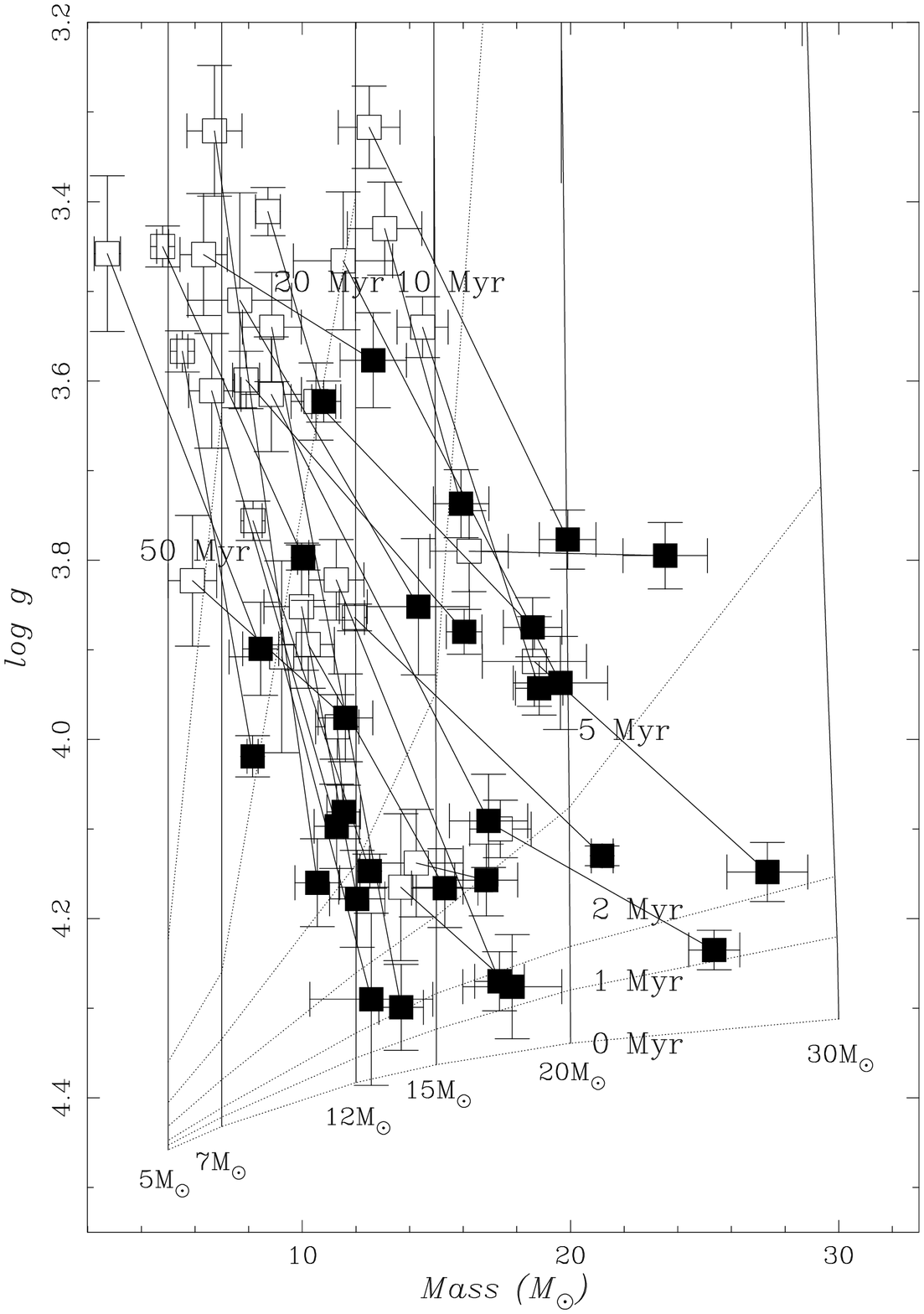} 
 \caption{The primary and secondary components (filled and open
symbols) of  28 SMC semi-detached systems and 1 SMC contact system plotted with 1-$\sigma$
uncertainties in the mass--surface-gravity plane (22 from this paper,
7 from H$^3$03). Lines join the primary and secondary of
each system. Also plotted are evolutionary tracks (solid lines) for
single stars of masses 5, 7, 12, 15, 20M$_{\odot}$ and isochrones
(dotted lines) for ages of 0, 1, 2, 5, 10, 20, 50Myr, taken from the
models of Girardi et al.\ (2000) for $Z=0.004$, appropriate for the
SMC.}  
\label{sdmlgg}
\end{figure*}

\begin{figure*}
\includegraphics[width=150mm]{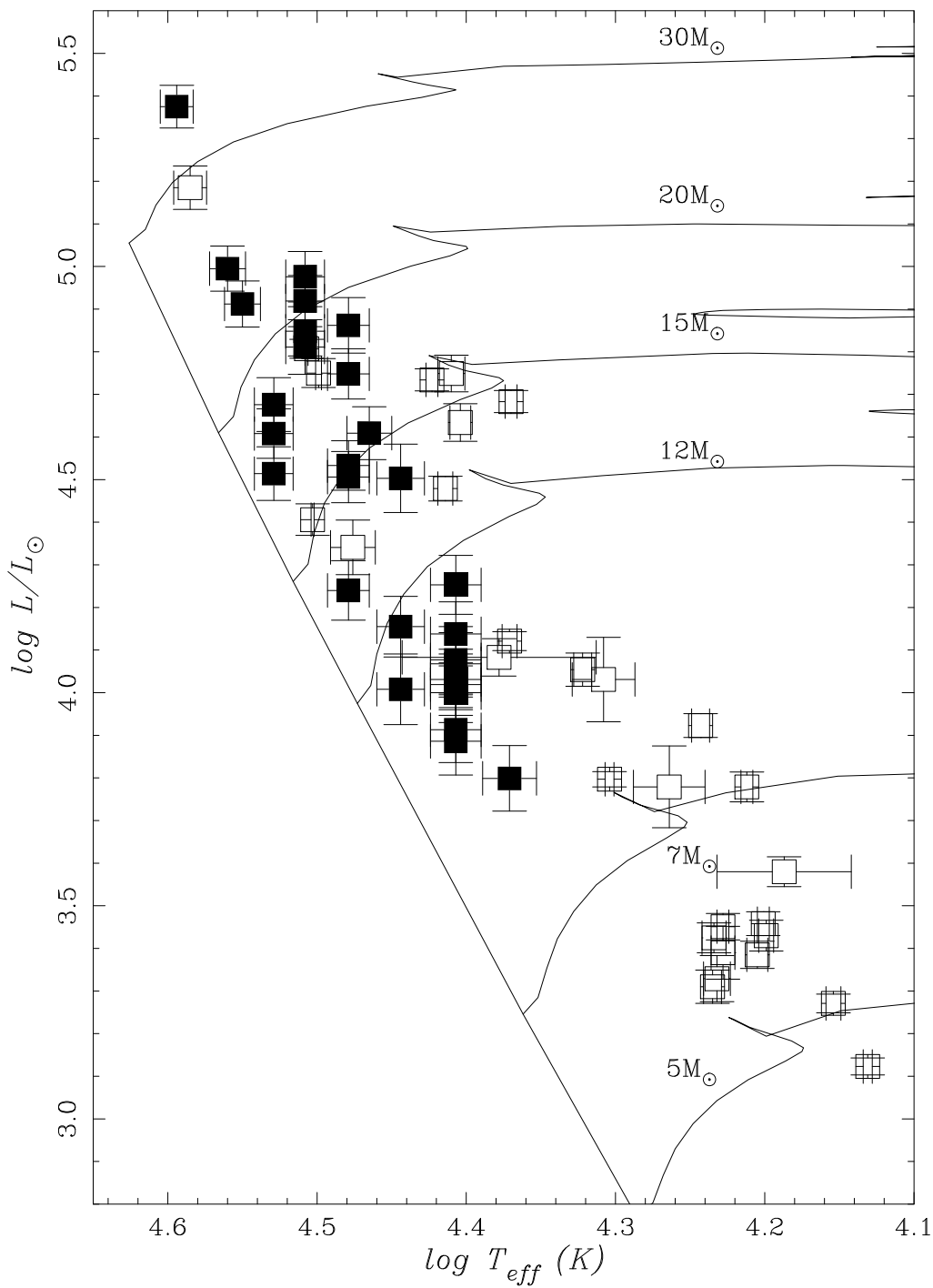} 
 \caption{The primary and secondary components (filled and open
 symbols) of  28 SMC semi-detached systems and 1 SMC contact system
 plotted with 1-$\sigma$ uncertainties in the $\log\,T_{\rm
 eff}$--$\log\,L/$L$_{\odot}$ plane (HR diagram) (22 from this paper,
 7 from H$^3$03).  Also plotted are the zero-age main sequence and
 evolutionary tracks for single stars of masses 5, 7, 12, 15, 20,
 30M$_{\odot}$, taken from the models of Girardi et al.\ (2000) for
 $Z=0.004$, appropriate for the SMC.}
\label{sdlgtlgl}
\end{figure*}

The distribution of the components of the 28 semi-detached systems in
the mass--surface-gravity plane (Fig.~\ref{sdmlgg}) could hardly be
more different from those of the detached systems. Note that the lines
connecting the components of each binary are all at obtuse angles
relative to the isochrones, in complete contrast to the detached
systems. In the HR diagram (Figure~\ref{sdlgtlgl}), the secondary
components appear to be systematically older than the primaries {\em
on the basis of single-star models,} and are all much larger than
expected for unevolved stars of the same mass. This is a very clear
signature that these stars are the mass-losing components seen in the
post-rapid-mass-transfer phase of case-A evolution (the longer-lived,
slow-mass-transfer phase of case~A). In our earlier discussion of
orbit and light-curve solutions for individual binaries, it was noted
that a substantial number of the semi-detached systems displayed clear
evidence of small depressions ($0.01-0.05$ mag) in their light curves
occurring just before the onset of primary eclipse. This evidence for
absorption by a weak mass-transfer stream (elaborated in
Section~\ref{X99121}) occurs in $\sim\,40\%$ of our semi-detached
sample, and may be taken as further evidence of these systems being in
the slow phase of mass transfer.

Whilst the locations of the observed secondary components of these
semi-detached systems in Figure~\ref{sdmlgg} and Figure~\ref{sdlgtlgl}
clearly show that they are evolved mass-losing stars, the observed
primary components occupy the main-sequence band. A natural question
is whether these stars, which are the mass gainers in case A
evolution, have the observed values of $log\,T_{\rm eff}$ and
$log\,L/L_{\odot}$ that are appropriate for single stars of the same
(newly-acquired) mass.  That is, does the evolutionary mass concur
with the observed dynamical mass also for the mass-gaining stars? We have
repeated the test carried out for the stars in detached systems on the
primary components of the semi-detached systems, and the individual
differences are listed in Table~\ref{sdmrgtl}.  The average difference
from Table~\ref{sdmrgtl} is $\Delta\,M=-1.0\pm4.4\,M_{\odot}$ (s.d.), which
becomes $\Delta\,M=-0.1\pm4.4\,M_{\odot}$ (s.d.) when the 7 primaries in
semi-detached systems from H$^3$03 are included.

This apparent agreement between evolutionary and observed masses
would suggest that the mass gainers do settle into equilibrium states
during the slow phase of case A with observed properties appropriate
for their newly-acquired masses. The examination by Braun \& Langer
(1995), which also reviewed earlier work on accretion on to massive
main-sequence stars, shows that the mass gainers may well adopt such
characteristics, provided that their core regions become again well
mixed with hydrogen and therefore have `rejuvenated' cores appropriate
for their new total mass. However, mass gainers may not always take
that route, but rather develop chemical structures unlike those of
single stars, where the core mass remains unchanged from before the
mass-transfer process. Such stars would appear to be under-luminous
for their newly-acquired mass. These observational data would suggest
that the mass gainers do have rejuvenated cores. However, the dispersion
about $\Delta\,M$ is a factor $\simeq\,2$ greater than that for the
detached systems, which suggests that the true answer could be more
complex.  

There are two extreme outliers in the distribution, the
primary components of 5~300549 (with $\delta\,M=-10.4\,M_{\odot}$) and
9~175323 (with $\delta\,M=+9.6\,M_{\odot}$), and if they are removed
from the sample $\Delta\,M=-0.1\pm3.6\,M_{\odot}$ (s.d.). [Systematic
errors in observed masses would lead to systematic errors in derived
distances and these two stars are respectively at distance moduli of
18.52 and 18.84, neither of which are unusual for the observed
distribution.] Perhaps the extra dispersion in $\Delta\,M$ is due to
some stars not having rejuvenated cores and therefore lower
luminosities, but the uncertainties of the data set are not
sufficiently small to resolve it. Perhaps one or two of the larger
positive $\delta\,M$ values could be due to binary systems being
observed at the end of the rapid phase of mass transfer where
evolution-model luminosities are larger than equilibrium values
(Wellstein, Langer \& Braun 2001). 

These results on SMC semi-detached binaries are similar to those for
semi-detached systems in the Milky Way galaxy (MWG) compiled by
Hilditch (2004). For that sample of only 10 systems, the average
difference is $\Delta\,M=+2\pm5\,M_{\odot}$, although some of those
discrepancies could be due to the adoption of different $T_{\rm eff}$
scales for O-type stars. This type of investigation should be pursued
further, with more data of higher quality and on a larger sample,
because it could help to resolve the still poorly known efficiency of
semi-convective mixing in mass-accreting stars (Wellstein et al. 
2001), and whether stellar cores do become rejuvenated in mass-gaining
events.

Regrettably, there are no published stellar-evolution models
for mass-exchanging binary systems in cases~A and~AB at SMC
metallicity ($Z \simeq 0.004$). However, the work of Wellstein
\& Langer (1999), and Wellstein, Langer \& Braun (2001) for 
solar-metallicity binaries ($Z=0.020$) provides many detailed examples
of what is expected for the evolution of close binaries with initial
orbital periods of a few days and initial masses in our observed
range.  Examination of their tracks for conservative case-A evolution,
followed by case~AB at later stages, shows sensible concurrence with
well-established semi-detached O,B binaries in the MWG, with the
observed secondaries identified as the mass-losing original primaries,
and the observed primaries as the mass-gaining original secondaries
(cf.\ Hilditch 2004).  The models indicate that initial orbital
periods of order 2--5 days remain around those values during the early
mass-reversal stages and into the slow phase of case~A, which is
expected to last of order 3--5 Myr before the case~AB phase starts
(when the orbital periods become much longer).

Our SMC semi-detached binaries fit this overall picture well.  Our
sample is defined by an upper orbital-period limit of $P\la5$ days,
and a faint absolute-magnitude limit of $M_{V}\simeq\,-3$,
corresponding to a spectral type on the main sequence of about B2.  Of
our 50 binary systems, 21 are in detached states and 28 are in
post-rapid-mass-transfer semi-detached states with properties that
seem to be consistent with conservative case-A evolutionary models.
The durations of the detached phases in the Wellstein \& Langer and
Wellstein, Langer \& Braun models are $\sim$3--12 Myr (dependent on
the initial orbital period and masses). The durations of the slow
phase of case~A are $\sim$1--5 Myr (again dependent on the particular
initial conditions). Note that the semi-detached phases of evolution
for some systems are as long as, or longer than, the detached phases
of others.

On the basis of these models, do we therefore expect an observed
non-coeval sample to display an approximately 3:4 division between
detached and semi-detached systems? Are the particular characteristics
of the stars consistent with {\em conservative} case-A evolution, or
is there an overall demand for some mass loss from the binary systems
to have occurred during the rapid mass-transfer phase that lasts only
some few $\times\,10^{4}$ years?  Binary-star evolutionary models at
SMC metallicity are required to compare with observational results on
the substantial sample of eclipsing binaries we have presented.  The
only contact binary in the sample is in deep contact, with a Mochnacki
(1984) fill-out factor $F=1.70$, and will presumably merge into a
single star of some 30M$_{\odot}$. Its origins are unclear, but it is
not a unique object amongst SMC or MWG OB binaries.

\section{Distances}
\label{Dists}

\begin{figure*}
\includegraphics[width=100mm,angle=270]{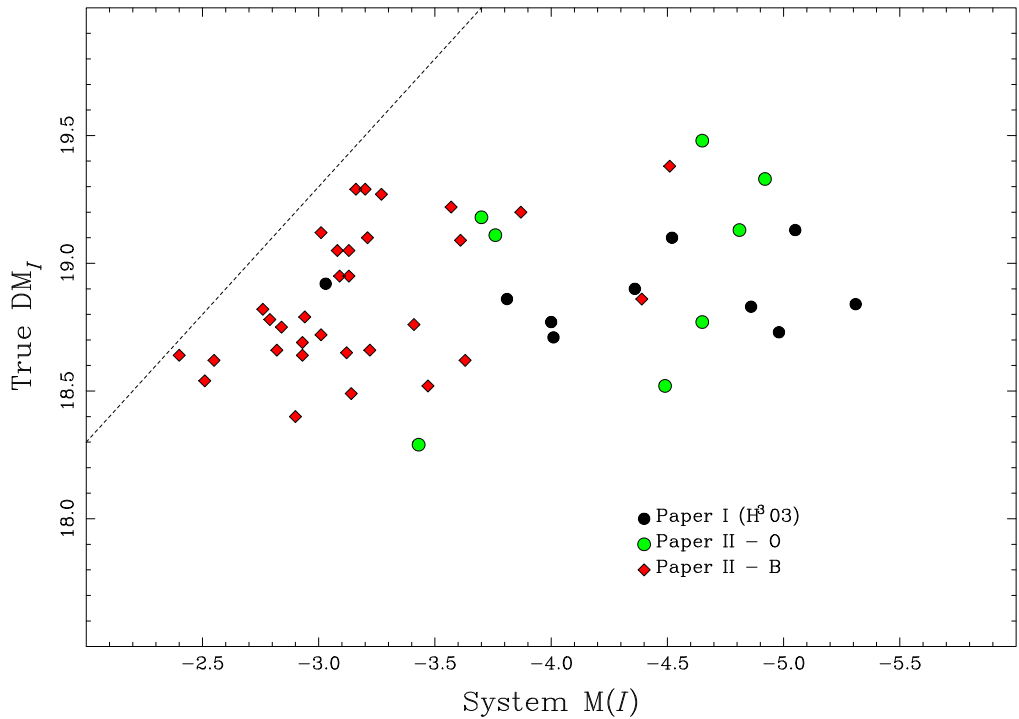} 
 \caption{Distance moduli as a function of absolute magnitude ($I$
 band).   O- and B-type stars are identified, as are results from
H$^3$03.   The diagonal line roughly corresponds to the magnitude
cut-off in our sample, and illustrates a modest selection bias.
}
\label{figDM}
\end{figure*}

Our methodology for distance determinations is fully described in
H$^3$03.  In brief, the photometric analysis gives the relative sizes
of the components in each system, and the spectroscopic orbit puts
these dimensions on an absolute scale.  An effective temperature is
assigned to the primary on the basis of its spectral type; its
absolute magnitude the follows from a light-curve synthesis code which
utilizes surface brightnesses from line-blanketed model atmospheres
(we use those given by Howarth \& Lynas-Gray 1989).  The light-curve
solution gives the relative brightnesses of the components, and hence
the absolute magnitude of the system.  The true distance modulus is
determined using $I$-band data, as the reddening is smaller, and the
intrinsic fluxes less temperature sensitive, at longer wavelengths;
extinction is determined from $(B-I)$ colours.  

Because many of our stars are somewhat cooler than the (predominantly
O-type) targets of Paper~I, a single 'hot star' colour is no longer
adequate; instead we adopted intrinsic colours from an approximate fit
to model results,
$$(B-I)_0 = \mbox{\rm{max}}[-0.6, 2.97 - 0.78\log(T_{\rm eff}(1))].$$
($log(T_{\rm eff})\ga\,4.3$).
In a handful of cases this leads to values for $A(I)$ that are smaller
than the expected foreground value of $\sim0.09$ ($E(B-V) \simeq
0.05$); we nonetheless retain the computed values in order not to bias
the mean distance modulus.

\begin{table*}
\caption[]{Distance determinations. $M(I)$ and $I_0$ are the absolute 
and observed $I$-band
quadrature magnitudes, and
DM is the true (reddening-free) distance modulus.   The $I$-band
extinction $A(I)$ is discussed in Section~\ref{Dists}.}
\label{dm_tab}
\begin{tabular}{lccccclcccc}
\hline

\multicolumn{1}{c}{OGLE}& 
$M(I)$ & $I_0$ & $A(I)$ &  DM & $\qquad\qquad\qquad$ & 
\multicolumn{1}{c}{OGLE}& 
$M(I)$ & $I_0$ & $A(I)$ &  DM \\
\hline
01 099121  & $-3.16$ & 16.30 & 0.17 & 19.29 &    &   06 221543  & $-3.61$ & 15.70 & 0.21 & 19.09     \\
04 056804  & $-3.22$ & 15.77 & 0.33 & 18.66 &    &   06 251047  & $-2.93$ & 15.97 & 0.21 & 18.69     \\
04 103706  & $-3.12$ & 15.85 & 0.31 & 18.65 &    &   06 311225 &  $-3.47$ & 15.20 & 0.14 & 18.52     \\
04 110409  & $-2.90$ & 15.87 & 0.36 & 18.40 &    &   06 319960 &  $-3.08$ & 16.09 & 0.11 & 19.05     \\
04 163552  & $-3.14$ & 15.41 & 0.06 & 18.49 &    &   07 066175 &  $-4.65$ & 14.46 & 0.33 & 18.77     \\
05 026631  & $-2.94$ & 15.99 & 0.14 & 18.79 &    &   07 120044 &  $-3.01$ & 15.81 & 0.10 & 18.72     \\
05 060548  & $-3.57$ & 15.86 & 0.21 & 19.22 &    &   07 142073 &  $-2.76$ & 16.20 & 0.14 & 18.82     \\
05 095194  & $-3.20$ & 16.23 & 0.14 & 19.29 &    &   07 189660 &  $-4.51$ & 15.06 & 0.19 & 19.38     \\
05 140701  & $-3.63$ & 15.24 & 0.25 & 18.62 &    &   07 193779 &  $-3.27$ & 16.19 & 0.19 & 19.27     \\
05 180064  & $-3.13$ & 16.11 & 0.19 & 19.05 &    &   07 255621 &  $-3.09$ & 16.13 & 0.27 & 18.95     \\
05 208049  & $-4.65$ & 14.98 & 0.15 & 19.48 &    &   08 087175 &  $-3.21$ & 16.04 & 0.14 & 19.10     \\
05 243188  & $-4.92$ & 14.64 & 0.23 & 19.33 &    &   08 104222 &  $-4.81$ & 14.56 & 0.24 & 19.13     \\
05 255984  & $-2.51$ & 16.24 & 0.20 & 18.54 &    &   08 209964 &  $-2.55$ & 16.24 & 0.17 & 18.62     \\
05 277080  & $-3.13$ & 16.07 & 0.25 & 18.95 &    &   09 010098 &  $-3.70$ & 15.59 & 0.11 & 19.18     \\
05 300549  & $-4.49$ & 14.16 & 0.14 & 18.52 &    &   09 047454 &  $-3.01$ & 16.26 & 0.15 & 19.12     \\
05 305884  & $-4.39$ & 14.65 & 0.18 & 18.86 &    &   09 064498 &  $-2.84$ & 16.03 & 0.12 & 18.75     \\
05 311566  & $-2.82$ & 16.03 & 0.19 & 18.66 &    &   10 037156 &  $-3.76$ & 15.49 & 0.14 & 19.11     \\
06 011141 &  $-3.87$ & 15.48 & 0.15 & 19.20 &    &   10 094559 &  $-2.40$ & 16.32 & 0.07 & 18.64     \\
06 152981  & $-2.93$ & 15.76 & 0.05 & 18.64 &    &   10 108086 &  $-3.41$ & 15.43 & 0.08 & 18.76     \\
06 180084 &  $-2.79$ & 16.17 & 0.18 & 18.78 &    &   10 110440 &  $-3.43$ & 15.03 & 0.16 & 18.29     \\
\hline
\end{tabular}
\end{table*}

Results are summarized
in Table~\ref{dm_tab}.
The mean true distance modulus for this sample is $18.88 \pm 0.30$
(s.d.), ranging 18.29--19.48, and the mean extinction is $A(I) = 0.179
\pm 0.070$ $(E(B-V) = A(I)/1.8 = 0.099 \pm 0.040)$.
The mean DM may be compared with the value of $18.89 \pm
0.14$ (range 18.71--19.13) obtained from  10 stars by H$^3$03.

In the present sample, there is no correlation of DM with $A(I)$,
$m(I)$, or $\gamma$ velocity, but there {\em
is} a modest correlation with $M(I)$ (Pearson and Spearman correlation
coefficients indicating that the null hypothesis of no correlation can
be rejected with $\sim$99.5\%\ confidence).  Fig.~\ref{figDM} shows
that this correlation arises from a simple selection effect: the
sample omits intrinsically faint systems with large derived DM, simply
because such systems are excluded by our apparent-magnitude cutoff.
This biases the mean DM towards a too-small value.  Restricting the
sample to $M(I)$ brighter than $-3.0$ removes any correlation and
gives a mean DM of $18.967 \pm 0.311$ from 29 stars, which we adopt as
the best estimate from the present sample.

Our results also show no statistically significant correlation of DM
with celestial co-ordinates.  However, Groenewegen (2000) has recently
analysed the intrinsic structure of the Magellanic Clouds, using
Cepheid photometry and adopting a simple triaxial model.  For the SMC,
he finds a $\sim$2kpc variation in mean distance ($\sim$0.07 in DM)
over the $\sim$2$^\circ$ spatial extent of our sample.  Our results
are consistent with this, a linear fit of distance along the projected
`axis' of the SMC suggesting that the NE extreme is $\sim$0.1m closer
than the SW.

Groenewegen also discusses the line-of-sight depth of the SMC, finding
an rms depth of 0.11m (3~kpc) for Cepheids.  While the distance moduli
from H$^3$03 and the present sample are consistent with a single mean
value, the present sample has a significantly larger dispersion (with
95\%\ confidence).  If the dispersion in the H$^3$03 sample (for which
internal and external error estimates were in close agreement) were
representative of the stochastic errors in the present sample, then
the residual dispersion of $\sim$0.28m($\sim$8kpc) in the present
sample would reflect astrophysical effects. However, this is surely a
substantial overestimate of the true dispersion in our parent
population, given the significantly larger observational uncertainties
in the current dataset.  We consider that our results are therefore
again consistent with the Cepheid studies in this respect.

Taking the weighted mean of the two eclipsing-binary samples (H$^3$03
and the present paper), our best estimate of the true distance modulus
to the SMC is $18.912 \pm 0.035$ ($60.6 \pm 1.0$~kpc), where the
quoted s.e. is, of course, an {\em internal} error; the result is
subject to systematic uncertainties at the $\sim$0.1~mag level, as
discussed in detail by H$^3$03.

Much work on the extragalactic distance scale is tied to the LMC
distance modulus (e.g., the HST Hubble Constant Key Project; Freedman
et al.\ 2001), because the structure of that galaxy has historically
been regarded as better understood.  We intend to address LMC binaries
directly in future work, but the difference in distance moduli,
SMC$-$LMC, is quite well documented as +0.5m (e.g., Udalski et al.\
1999, Groenewegen 2000).  Our SMC results therefore imply an LMC DM of
$18.41 \pm 0.04 \pm 0.1$, as compared with a value of 18.50 adopted in
the Key Project work.

\section{Conclusion}

Together with results from H$^3$03, our sample of SMC binaries
includes 50 systems.  With only two or three obvious exceptions, the
typical uncertainties are $\pm$10\%\ on masses, $\pm$4\%\ on radii,
and $\pm$0.07 on $\log\,L$ -- remarkably good values
considering the limited spectral resolution and data quality available
for this study, and the use only of single-passband light curves.  
For comparison, in a recent review of the observed properties of
high-mass eclipsing binaries, Hilditch (2004) adopted similar upper
limits for the uncertainties on these quantities derived for Galactic
O,B binaries, and found only 22 systems that satisfied these criteria.
The aims of this project have thus clearly been realised, in that we
have provided good determinations of fundamental parameters for the
largest sample of O,B type eclipsing binaries achieved for any galaxy
(including the Milky Way), and have made a direct determination of the
SMC distance modulus that has exceptionally high precision (and very
good accuracy).   We find good agreement between our observations
and the evolutionary models, with no evidence for a `mass discrepancy'
in the models.

\section{Acknowledgments}

This paper is based largely on data obtained at the Anglo-Australian
Observatory, and we thank Dr Terry Bridges, in particular, for
obtaining the service observations for us in 2001 and 2003, and for
providing much helpful advice in the \textsc{2dFDR} data-reduction
procedures. We also thank Dr Elizabeth Corbett for obtaining the 2002
data.  We are grateful to the OGLE team for making their excellent
photometric database publicly available, and to Dr Graham Hill for the
use of his \textsc{light2} code. This research benefitted from a PPARC
short-term visitor grant, PPA/V/S/2001/00550, awarded to the
Department of Physics, University of Exeter.

\end{document}